\documentclass{aa}
\usepackage{txfonts}
\usepackage{graphicx}
\usepackage{subcaption}
\usepackage[colorlinks=true,allcolors=blue]{hyperref}

\usepackage{gensymb}

\begin{document} 

\authorrunning{Mountrichas et al.}
\titlerunning{The Incidence of X-ray AGN and Non-AGN Galaxies in the Far-Infrared}


\title{The incidence of X-ray AGN and non-AGN galaxies in the far-infrared: insights into host galaxy properties and AGN obscuration}

\author{G. Mountrichas\inst{1}, F. J. Carrera\inst{1}, I. Georgantopoulos\inst{2}, S. Mateos\inst{1}, A. Ruiz\inst{2}, A. Corral\inst{1}}
          
     \institute {Instituto de Fisica de Cantabria (CSIC-Universidad de Cantabria), Avenida de los Castros, 39005 Santander, Spain
              \email{gmountrichas@gmail.com}
           \and
             National Observatory of Athens, Institute for Astronomy, Astrophysics, Space Applications and Remote Sensing, Ioannou Metaxa
and Vasileos Pavlou GR-15236, Athens, Greece       
}

\abstract{We investigate the far-infrared (far-IR) incidence of X-ray-selected active galactic nuclei (AGN) and galaxies that do not host an AGN (non-AGN) as a function of stellar mass (M$_*$), star formation rate (SFR), and specific black hole accretion rate ($\lambda_{\text{sBHAR}}$), using data from five well-characterized extragalactic fields (COSMOS, XMM-LSS, Stripe82, ELAIS-S1, and CDFS-SWIRE). We construct spectral energy distributions (SEDs) using optical-to-far-IR photometry to derive host galaxy properties and assess AGN obscuration, while X-ray absorption is quantified using the 4XMM-DR11s catalog. Our final sample comprises 172\,697 non-AGN galaxies (53\% Herschel-detected) and 2\,417 X-ray AGN (73\% Herschel-detected), with $\rm 10 < log\,[M_*/M_\sun] < 12$ and $\rm 0 < z < 2$. We find that X-ray AGN exhibit a relatively flat far-IR detection rate across stellar mass and specific SFR ($\rm sSFR = SFR / M_*$), unlike non-AGN galaxies, where detection correlates strongly with star formation. Far-IR detection among AGN decreases with increasing $\lambda_{\text{sBHAR}}$, even as their SFR rises. Our results suggest that X-ray AGN are preferentially found in gas-rich environments, where both star formation and black hole accretion are fueled by the presence of cold gas. The far-IR
incidence of X-ray AGN remains high across all sSFR bins, indicating that these AGN can coexist with ongoing star formation for extended periods, in line with a scenario where AGN feedback regulates rather than abruptly quenches star formation. We also find that comparing AGN and non-AGN SFRs without separating Herschel-detected from non-detected sources introduces biases. Obscured AGN show ~10\% higher far-IR detection rates than unobscured ones, yet at similar $\lambda_{\text{sBHAR}}$, unobscured AGN tend to have higher SFR. This may indicate that obscured AGN reside in dustier environments where moderate star formation still contributes to far-IR emission. Our results support a scenario where AGN and star formation coexist in gas-rich galaxies, with AGN feedback acting as a regulatory process over extended timescales and not necessarily quenching.}

\keywords{}
   
\maketitle  

\section{Introduction}

    Understanding the relationship between active galactic nuclei (AGN) and their host galaxies is crucial for unraveling the co-evolution of galaxies and supermassive black holes (SMBHs). AGN offer valuable insights into the accretion processes of SMBHs and the physical conditions of the surrounding interstellar medium (ISM). A key question in galaxy evolution is the role of AGN in regulating star formation, potentially quenching it through feedback processes.

    Numerous studies have explored the connection between AGN activity, as indicated by X-ray luminosity (L$_\mathrm{X}$), and the star formation rates (SFR) of their host galaxies \citep[e.g.,][]{Lutz2010, Page2012, Harrison2012, Rosario2012, Santini2012, Rovilos2012, Shimizu2015, Rosario2013, Mullaney2015, Masoura2018, Bernhard2019, Florez2020, Pouliasis2022}. More recently, comprehensive analyses by \citet{Mountrichas2021b, Mountrichas2022a, Mountrichas2022b, Mountrichas2024c} compared the SFR of AGN with that of galaxies without an active SMBH. Their findings indicate that X-ray AGN exhibit similar SFR to galaxies without AGN up to a certain L$_\mathrm{X}$ threshold (43-44 erg/s in logarithmic scale), beyond which AGN hosts show elevated SFR. This threshold appears to increase with the stellar mass (M$_*$) of the host. \citet{Cristello2024} further showed that at high M$_*$, comparisons between AGN and non-AGN become challenging due to the turnover of the galaxy main sequence, and that SFR–L$_\mathrm{X}$ trends are largely driven by the underlying SFR–M$_*$ relation.

    In addition, recent studies have emphasized the importance of the specific black hole accretion rate \citep[$\lambda_{\text{sBHAR}}$, e.g.,][]{Georgakakis2017a, Georgakakis2017b, Aird2019} as a critical parameter for understanding the interaction between AGN activity and star formation. This parameter, which measures the accretion efficiency relative to the Eddington limit, has been shown to correlate with both L$_\mathrm{X}$ and the star formation efficiency of AGN host galaxies \citep[e.g.,][]{Aird2019, Torbaniuk2021, Mountrichas2023d, Torbaniuk2024, Guetzoyan2025}.

     Obscuration is a key aspect of AGN classification and evolution, typically attributed to an optically thick, dusty torus surrounding the central engine \citep[e.g.,][]{Urry1995}. This torus can obscure the AGN in certain viewing angles, giving rise to the observed distinction between type 1 and type 2 AGN. However, there is growing evidence that obscuration may also arise on larger, host-galaxy scales, particularly in systems with high star formation and dust content \citep[e.g.,][and references therein]{Ricci2017, Almeida2017}. Disentangling the relative contributions of torus-scale versus host-scale dust is essential for understanding the diversity of AGN properties and their connection to galaxy evolution.

    Importantly, the origin of far-IR emission in luminous AGN, particularly QSOs, remains an active and controversial topic. Some studies \citep[e.g.,][]{Symeonidis2016, Symeonidis2022} argue that a significant fraction of the far-IR emission beyond 100\,$\mu$m, can arise from AGN-heated dust rather than host-galaxy star formation. Others challenge this view, attributing most of the far-IR output to dusty star-forming regions in the host \citep[e.g.,][]{Stanley2015, Mullaney2015}. Clarifying this issue is essential for understanding both AGN energetics and host galaxy evolution.

    Examining the incidence of AGN across different wavelengths provides critical insights into the physical mechanisms linking black hole growth to host galaxy properties. For instance, \citet{CastelloMor2013} explored obscuration trends in X-ray AGN with Spitzer/IRAC detections, while \citet{Igo2024} linked radio AGN to host properties in the eFEDS fields. \citet{Zhang2025} showed that the low star-forming fraction of radio AGN is primarily driven by their preference for massive hosts, and \citet{Kondapally2025} found that HERGs reside in star-forming systems while LERGs span a broader SFR range—highlighting cold versus hot gas accretion pathways.

    This study aims to clarify a fundamental but still unresolved issue: how common far-infrared (far-IR) emission is among X-ray-selected AGN, and how that compares to non-AGN galaxies, across a broad range of host and AGN properties. While prior works have investigated the SFRs of AGN hosts using Herschel data, a comprehensive analysis of far-IR incidence—its dependence on stellar mass, redshift, black hole accretion rate, and obscuration—has not yet been fully explored. By focusing not on total far-IR luminosity but on detection rates, our analysis provides a complementary approach: it examines whether AGN are systematically found in dust-rich, star-forming environments more often than matched non-AGN galaxies. We also examine how AGN obscuration and X-ray absorption relate to far-IR detection, providing new observational constraints on the dust and gas content of AGN hosts. For that purpose, we use data from five well-characterized extragalactic fields, namely COSMOS, XMM-LSS, Stripe82, ELAIS-S1, and CDFS-SWIRE. We construct and fit spectral energy distributions (SEDs) from optical to far-IR, and cross-match with the 4XMM-DR11s catalog to identify AGN. We classify them as type 1 or 2 based on inclination angles inferred from SED fitting and also group them by X-ray absorption levels based on hydrogen column density (N$_\mathrm{H}$) measurements.

    This paper is organized as follows: In Sect. \ref{sec_data}, we describe the datasets and AGN selection methodology. In Sect. \ref{sec_analysis}, we detail our SED fitting procedure, completeness criteria, and final sample construction. Sect. \ref{sec_fraction_all} analyzes far-IR detection rates as functions of M$_*$, SFR, $\lambda{\text{sBHAR}}$, and redshift. Sect. \ref{sec_sfr} compares the SFRs of AGN and non-AGN galaxies across detection status, and Sect. \ref{sec_agn_type} explores the role of obscuration and absorption. Finally, Sect. \ref{sec_conclusions} summarizes our results and conclusions.

\section{Data}
\label{sec_data}

In this section, we describe the five fields from which our galaxy datasets are drawn and outline the criteria we implement to ensure sufficient photometric coverage for conducting a robust SED fitting analysis. We also describe how we identify X-ray detected AGN among the galaxy sources.

\subsection{The galaxy catalogues}
To compile our galaxy reference sample, we utilize publicly available catalogs from the \textsl{Herschel} Extragalactic Legacy Project (HELP\footnote{\href{http://hedam.lam.fr/HELP/dataproducts/}{HELP Data Products}}) \citep[][]{Shirley2019, Shirley2021}. This project compiles, curates, and standardizes data, creating derived products for 23 of the most significant multiwavelength extragalactic survey fields. The dataset comprises approximately 179 million objects, covering 1\,270 deg$^2$, defined by the \textsl{Herschel} Multi-tiered Extragalactic Survey (HerMES) and the \textsl{Herschel} Atlas survey (H-ATLAS). The catalogues include photometry from optical, near-infrared, mid-infrared, and far-infrared bands, obtained through positional cross-matching of 51 public surveys. Data Release 1 (DR1) features a comprehensive map of the largest SPIRE extragalactic field, spanning 385 deg$^2$, and includes 18 million measurements of PACS and SPIRE fluxes. Photometric redshifts are also provided, utilizing the methodology presented by \cite{Duncan2018a, Duncan2018b}, which combines multiple template-fitting and machine-learning techniques to produce a consensus estimate with well-calibrated uncertainties. Table 3 in \cite{Shirley2021} details the specifics of each field included in HELP, such as the number of sources, coverage area, and the availability of photometric and spectroscopic redshifts. Below, we briefly summarize the five fields utilized in our study.

\subsubsection{COSMOS}

The Cosmic Evolution Survey (COSMOS) field covers 2 deg$^2$ of the sky with central coordinates of $\alpha = 10^\mathrm{h}00^\mathrm{m}28.6^\mathrm{s}, \delta = +02^\circ12'21.0''$ \citep{Scoville2007}, and it is one of the most comprehensively studied extragalactic fields, with multiwavelength data spanning from X-ray to radio wavelengths. The field includes deep optical and near-infrared observations from the Hubble Space Telescope (HST), as well as data from ground-based telescopes and space missions such as \textit{Spitzer}, \textsl{Herschel}, and \textit{XMM-Newton} \citep{Scoville2007, Laigle2016}. In the COSMOS field, we employ the data used in \cite{Mountrichas2022a}. There are about $2.5$ million galaxies in the COSMOS field as included in the HELP database. Around 500\,000 of these are in the UltraVISTA survey \citep[see also][]{Laigle2016}
which covers 1.38~deg$^2$ of the COSMOS field, after removing the masked objects \citep[see Fig. 1 in ][]{Laigle2016}. This survey has deep near-infrared observations (J, H, Ks photometric bands)
that allow us to derive more accurate host galaxy properties through SED fitting (see below).

\subsubsection{Stripe82}

Stripe82 is a region of the sky located along the celestial equator, covering about 300 deg$^2$, with right ascension ranging from 20h to 04h and a declination of $\pm$1.25°. It is part of the Sloan Digital Sky Survey (SDSS) and benefits from deep, repeated imaging  \citep{Jiang2014}. The repeated scans of Stripe82 have resulted in deep co-added images that reach $2-3$ magnitudes fainter than the single-epoch SDSS data, depending on the band, making it possible to detect faint galaxies and study the low surface brightness universe \citep{Annis2014}. Additionally, Stripe82 has been observed across multiple wavelengths, including near-infrared data from the UKIRT Infrared Deep Sky Survey (UKIDSS) and mid-infrared data from the \textit{Spitzer} Space Telescope. We note, that due to the size of the Stripe82 field, we have restricted the galaxy sources to only those regions of the field that overlap with the 4XMMDR11 detections (see below).

\subsubsection{XMM-LSS}

The XMM-Newton Large Scale Structure (XMM-LSS) field covers an area of approximately 11 deg$^2$, centered near a right ascension of 02h 20m and a declination of -04° \citep{Pierre2004, Chiappetti2005}. The field has been extensively observed in X-rays by \textit{XMM-Newton} and  benefits from deep multiwavelength data, including optical imaging from the Canada-France-Hawaii Telescope Legacy Survey \citep[CFHTLS;][]{Gwyn2012} and near-infrared data from the Visible and Infrared Survey Telescope for Astronomy \citep[VISTA;][]{Jarvis2013}.

\subsubsection{ELAIS-S1}

The European Large Area ISO Survey - South 1 (ELAIS-S1) field is a deep survey centered at a right ascension of 00h 37m and a declination of -44°, covering approximately 4 deg$^2$. It was initially observed in the mid-infrared by the Infrared Space Observatory (ISO) as part of the European Large Area ISO Survey (ELAIS) \citep{Oliver2000}. ELAIS-S1 has since been complemented by extensive multiwavelength data, including optical, near-infrared, and far-infrared observations. 

\subsubsection{CDFS-SWIRE}

The Chandra Deep Field South - Spitzer Wide-area Infrared Extragalactic Survey (CDFS-SWIRE) field is an extensively studied region centered at a right ascension of 03h 32m and a declination of -28°, covering approximately 8 deg$^2$. It encompasses multiwavelength coverage, including infrared data from the \textsl{Spitzer} Wide-area Infrared Extragalactic Survey (SWIRE) \citep{Giacconi2001, Lonsdale2003}.

\subsection{Photometric selection criteria}

For our analysis, it is essential to obtain reliable estimates of galaxy properties through SED fitting. To achieve this, we require all galaxies to have detections in the following photometric bands: $u, g, r, i, z, J, H, K_s$, IRAC1, IRAC2, and MIPS/24. The IRAC1, IRAC2, and MIPS/24 bands correspond to the 3.6\,$\mu$m, 4.5\,$\mu$m, and 24\,$\mu$m photometric bands of Spitzer. A total of 682\,889 galaxies across the five fields meet these photometric criteria, allowing us to perform SED fitting analysis on these sources. Specifically, 295\,266 galaxies are in the COSMOS field, 320\,319 in Stripe82, 24\,353 in XMM-LSS, 19\,840 in ELAIS-S1, and 23\,111 in CDFS-SWIRE.

\begin{figure}
\centering
  \includegraphics[width=0.75\columnwidth, height=4.cm]{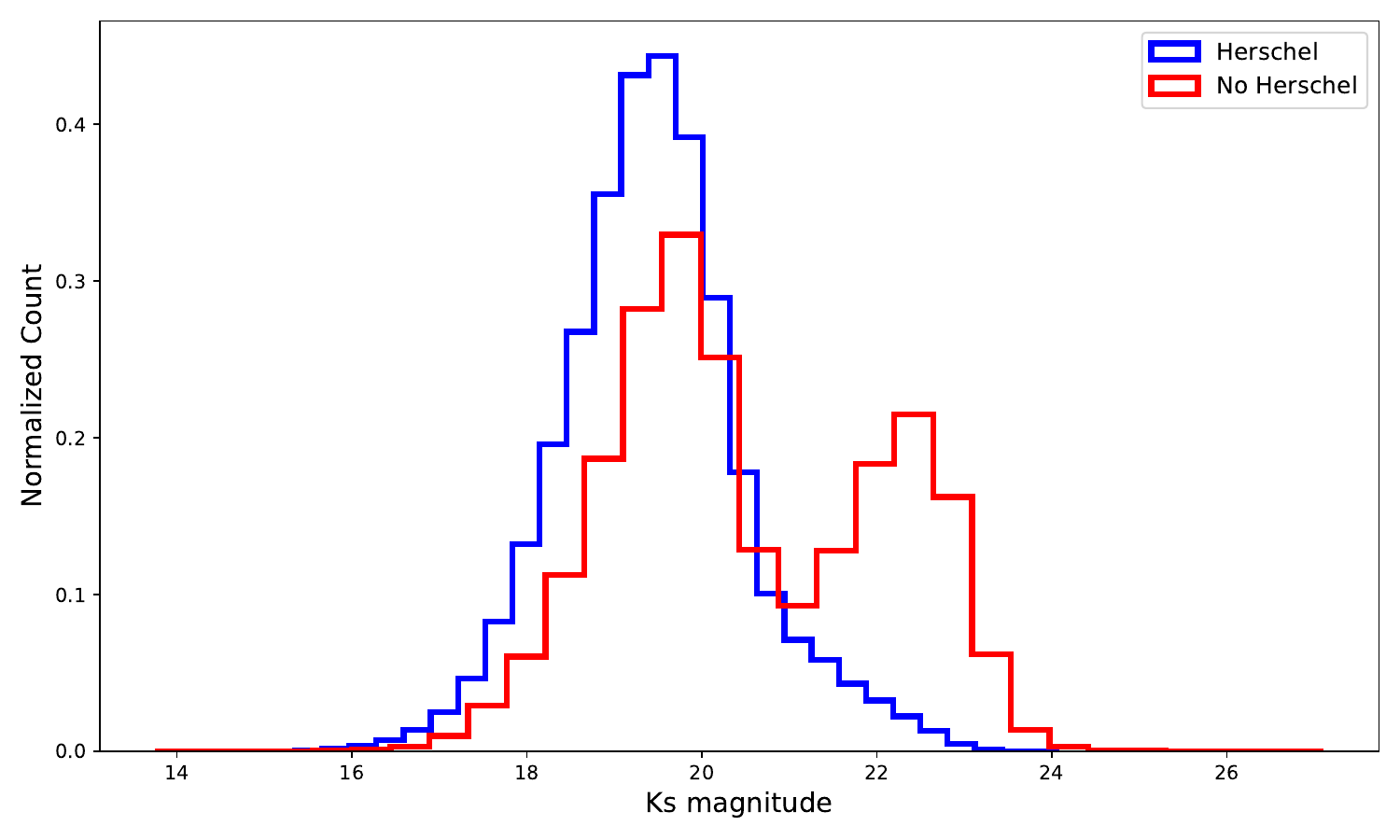} 
  \caption{Distribution of $\rm K_s$ for 
  \textsl{Herschel} detected and non-detected sources. The sample shown here corresponds to the dataset defined in Section~\ref{sec_herschel_crit}, before the application of the $\rm K_s$ magnitude cut and the stellar mass limits used to define the final sample.}
  \label{fig_ks_4datasets}
\end{figure}  

\begin{figure}
\centering
  \includegraphics[width=0.99\columnwidth, height=4.5cm]{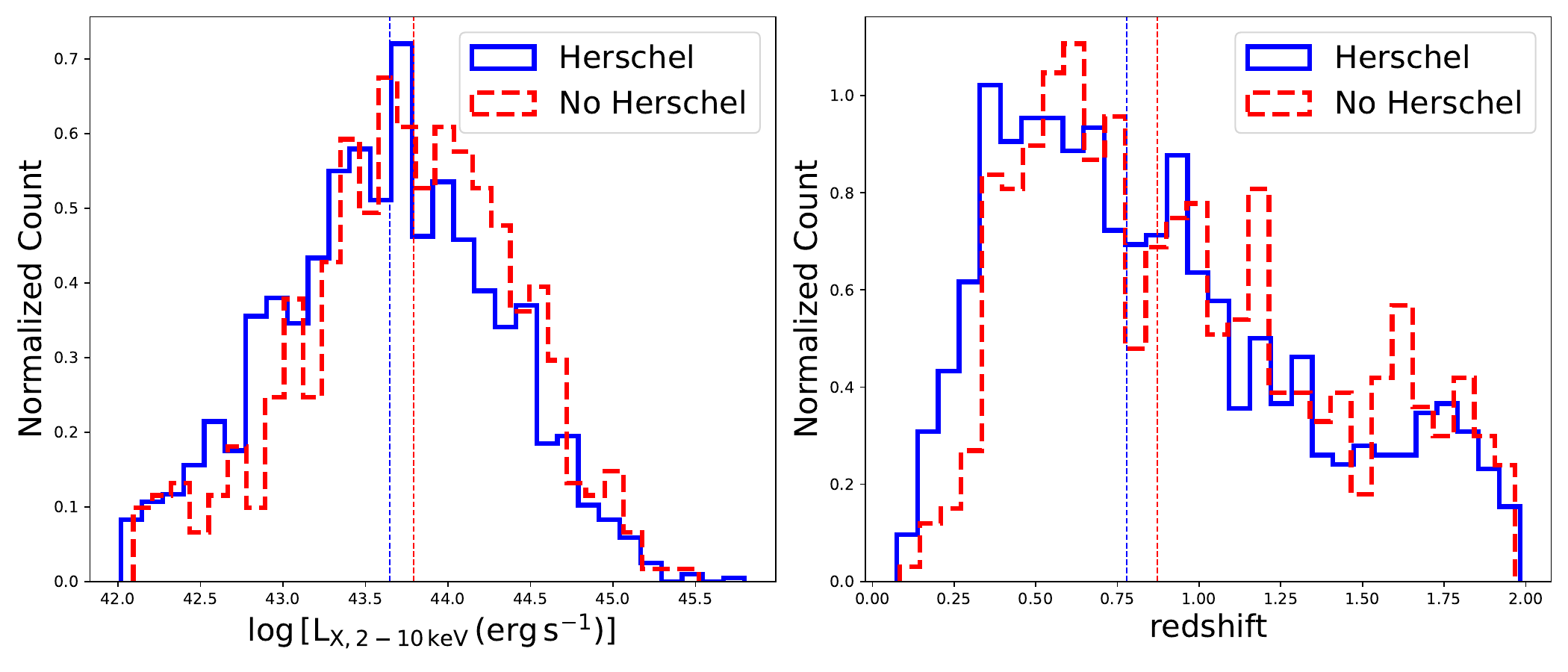} 
  \includegraphics[width=0.99\columnwidth, height=4.5cm]{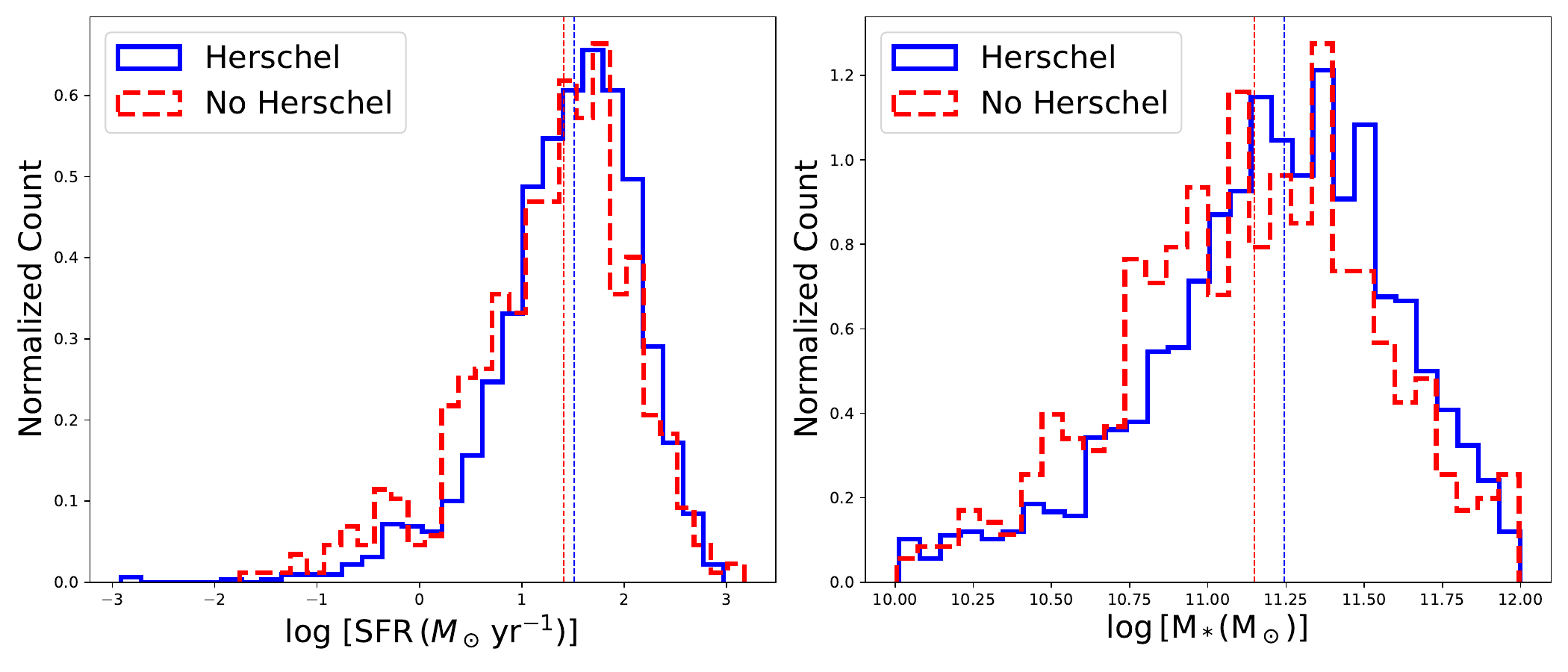} 
  \caption{Distributions of $\rm log\,[L_{X,2-10keV}(ergs^{-1})]$ (top, left panel), redshift (top, right panel), SFR (left, bottom panel) and M$_*$ (right, bottom, panel) for X-ray detected AGN with and without far-IR photometric detection. Vertical lines correspond to the median values of the two subsets.}
  \label{fig_main_prop_agn}
\end{figure}  

\begin{table*}[h!]
\centering
\setlength{\tabcolsep}{2.pt} 
\caption{Median values of the main properties of the X-ray AGN and no-X-ray galaxies samples, with and without \textsl{Herschel} detection.}
\begin{tabular}{cccccc}
\hline
sample & no. of sources & $\rm log\,[L_{X,2-10keV}(ergs^{-1})]$ & $\rm log\,[SFR (M_\odot yr^{-1})]$ & $\rm log\,[M_*(M_\odot)]$ & redshift  \\
\hline
X-ray AGN (\textsl{Herschel})  &  1\,782 &  43.68 & 1.56 & 11.25 &   0.76\\
X-ray AGN (no \textsl{Herschel}) &  659 & 43.80 & 1.42 & 11.15 &   0.87\\
no-X-ray detected galaxies (\textsl{Herschel}) & 91\,537 & - & 1.05 & 10.90 & 0.46  \\
no-X-ray detected galaxies (no \textsl{Herschel}) & 81\,160 & - & 0.32 & 10.84 & 0.50  \\
\end{tabular}
\label{table_main_properties}
\end{table*}

\begin{figure} 
\centering
  \includegraphics[width=0.7\columnwidth, height=4.cm]{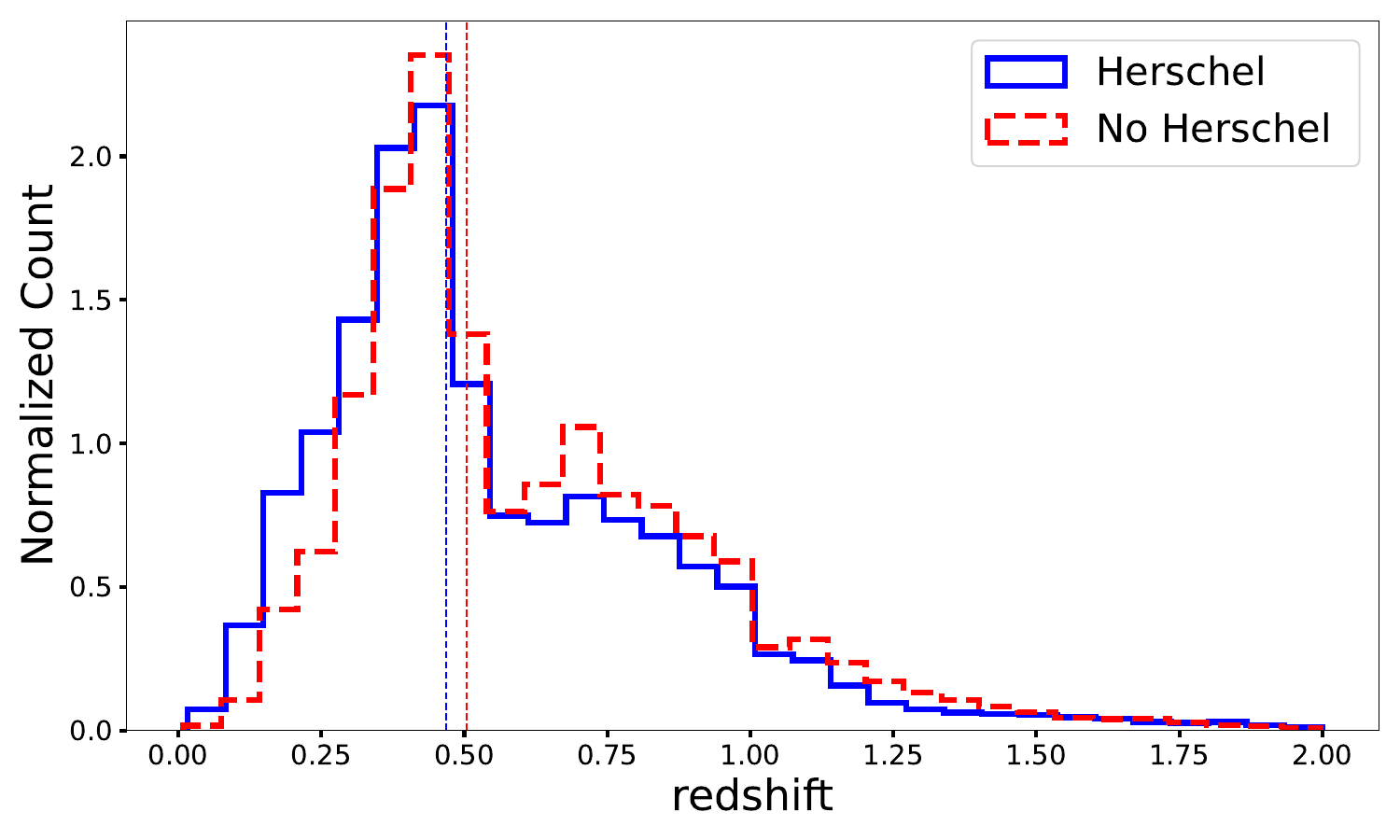}
  \includegraphics[width=0.7\columnwidth, height=4.cm]{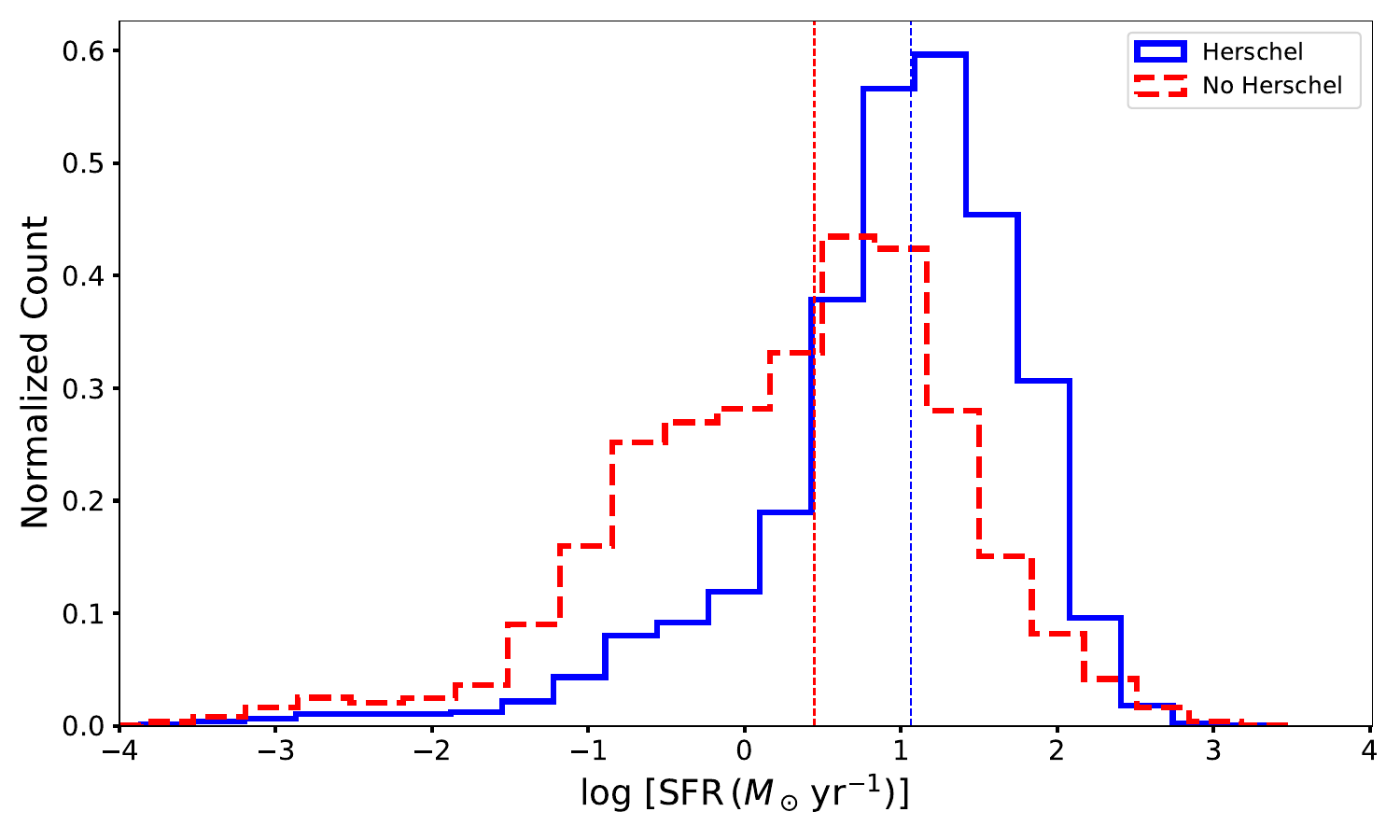}   
  \includegraphics[width=0.7\columnwidth, height=4.cm]{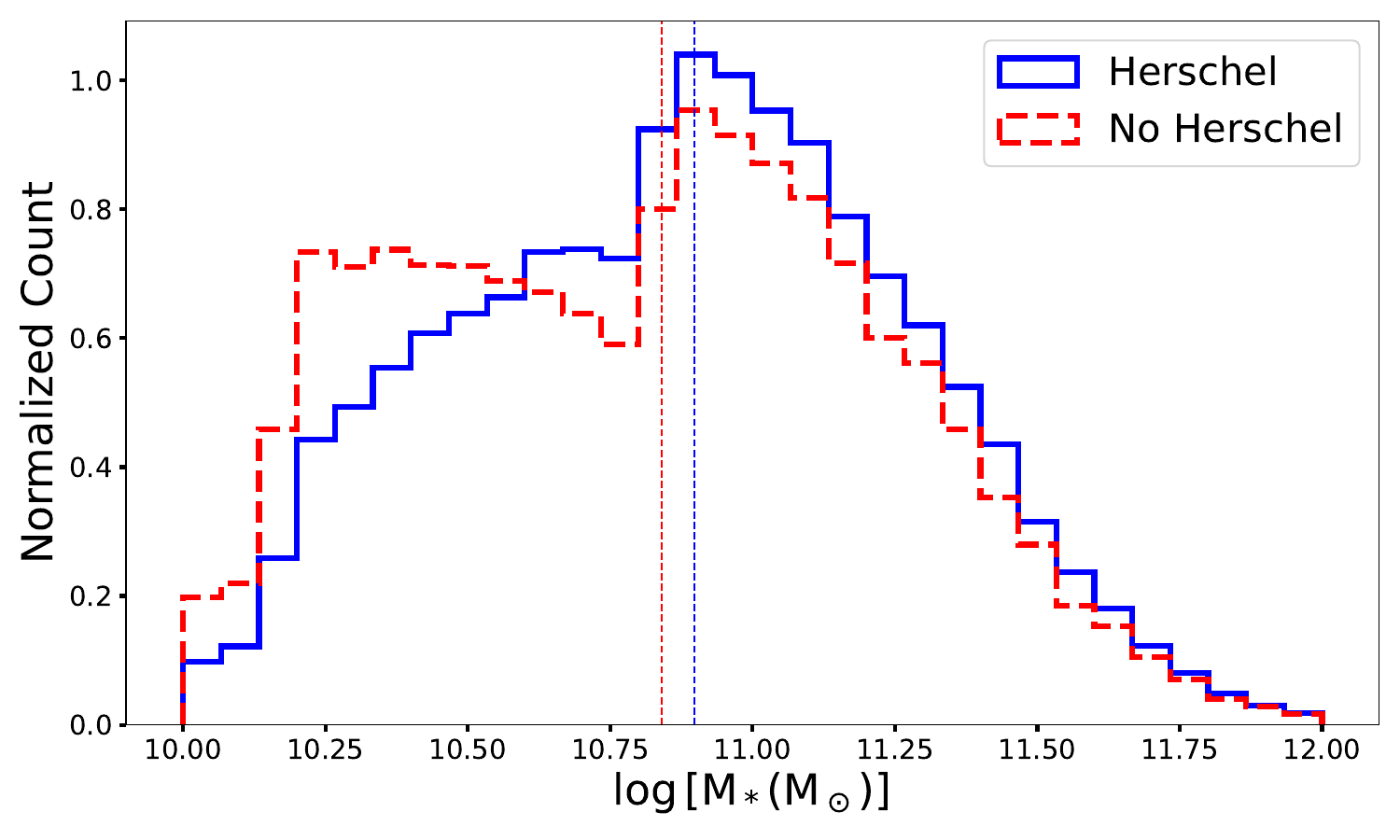} 
  \caption{Distributions of redshift (top panel), SFR  (middle panel) and M$_*$ (bottom panel) for no-X-ray detected galaxies with and without far-IR photometric detection. Vertical lines correspond to the median values of the two subsets.}
  \label{fig_main_prop_gals}
\end{figure}  

\subsection{X-ray sources}
\label{sec_xray_sources}

To identify X-ray AGN among the 682\,889 galaxies in the five fields described above, we use the 
4XMM-DR11s catalog \citep{Webb2020, Traulsen2020}. 4XMM-DR11s is the eleventh data release from the XMM-Newton Serendipitous Source Catalogue, an extensive collection of X-ray sources detected by the European Space Agency's (ESA) XMM-Newton observatory. Released in 2021, the catalog includes data from observations performed between 2000 and 2020, covering over two decades of X-ray observations. It comprises 1\,488 stacks (or groups), with most stacks consisting of two observations, while the largest stack includes 357 observations. The catalogue contains a total of 358\,809 sources, of which 275\,440 have multiple contributing observations. The main advantage of stacking observations is that the repeatedly-observed source parameters are determined with higher precision, also improving the study of their long term variability. The 4XMM-DR11s catalogue achieves a depth of approximately $3.3 \times 10^{-15}$ erg/cm$^2$/s in the soft X-ray band (0.2-2 keV) and $1.1 \times 10^{-14}$ erg/cm$^2$/s in the hard X-ray band (2-12 keV).

The cross-matching of the X-ray dataset with the galaxy sample was conducted using the xmatch tool from the astromatch package\footnote{https://github.com/ruizca/astromatch}. This tool enables the matching of multiple catalogs and provides Bayesian probabilities for associations or non-associations using the astrometry of the sources \citep{Pineau2017, Ruiz2018}. We retained only those sources with a high association probability, specifically those exceeding 90\%. In instances where a source was linked to multiple counterparts, we selected the association with the highest probability. This process identified 4\,589 X-ray sources within the HELP catalogues. These sources were excluded from the final galaxy dataset, which we refer to as the non-X-ray detected galaxy sample. To identify AGN among the X-ray sources, we applied a luminosity threshold of \(\rm log\,[L_{X,2-10keV}(ergs^{-1})]>42\). The L$_X$ refers to the intrinsic (i.e. absorption-corrected) 2–10 keV X-ray luminosity that was derived through spectral fitting using an absorbed powerlaw model, based on the posterior distributions of flux and photon index obtained from the X-ray spectral analysis \citep{Viitanen2025}. This yielded a total of 4\,437 X-ray AGN. We note that this L$_X$ threshold is commonly adopted to ensure that the X-ray emission originates from an active nucleus rather than stellar processes such as X-ray binaries \citep[e.g.,][]{Georgakakis2008, Aird2012, Mountrichas2024b}. Raising the threshold to \(\rm log\,[L_{X,2-10keV}(ergs^{-1})]>42\).5 would exclude fewer than 3\% of our AGN sample and would not alter our results, as the associated changes lie well within our statistical uncertainties.

\section{Analysis}
\label{sec_analysis}

\subsection{Calculation of galaxy properties}
\label{sec_cigale}

We derived galaxy properties such as SFR and M$_*$ through SED fitting using the CIGALE code \citep{Boquien2019, Yang2020, Yang2022}, adopting the same templates and parameter grid as in our previous studies \citep[e.g.,][]{Mountrichas2021c, Mountrichas2022a, Mountrichas2022b, Mountrichas2023d, Mountrichas2024c}. Details of the models and parameter space are provided in Appendix~\ref{appendix_sed}, while the quality criteria used to select sources with reliable SED fitting results are described in Appendix~\ref{appendix_sed_criteria}. A total of 3\,565 X-ray detected AGN and 398\,439 non-X-ray sources satisfy these criteria.

The reliability of the SFR measurements, both in the case of AGN and non-AGN systems, has been examined in detail in our previous works and, in particular, in Sect. 3.2.2 in \cite{Mountrichas2022a}. Finally, we note that the AGN module is used when fitting the SEDs of non-X-ray detected galaxies. This allows us to identify AGN that are not detected in X-rays \citep[e.g.,][]{Pouliasis2020} and exclude them from the galaxy control sample. Approximately 4\% of the non-X-ray-detected galaxies are excluded based on this criterion (see Sect.~\ref{appendix_SEDAGN}). The remaining sources—i.e., galaxies not detected in X-rays and without significant AGN contribution in their SEDs—are referred to as non-AGN galaxies throughout the rest of the paper.

\subsection{Mass completeness}
\label{sec_mass_complet}

We define complete X-ray AGN and non-AGN samples by applying mass-completeness limits, which are essential to avoid potential selection biases in our analysis. To achieve this, we calculate the mass completeness in four redshift bins within \(0 < z < 2\), with a step of 0.5, following the method described in \cite{Pozzetti2010}. The details and the mass completeness limits in each one of the five fields are shown in Appendix \ref{appendix_mass_complete}. There are 228\,152 non-AGN galaxies and 3\,046 X-ray AGN that meet the mass completeness limits.

\begin{figure}
\centering
  \includegraphics[width=0.99\columnwidth, height=9cm]{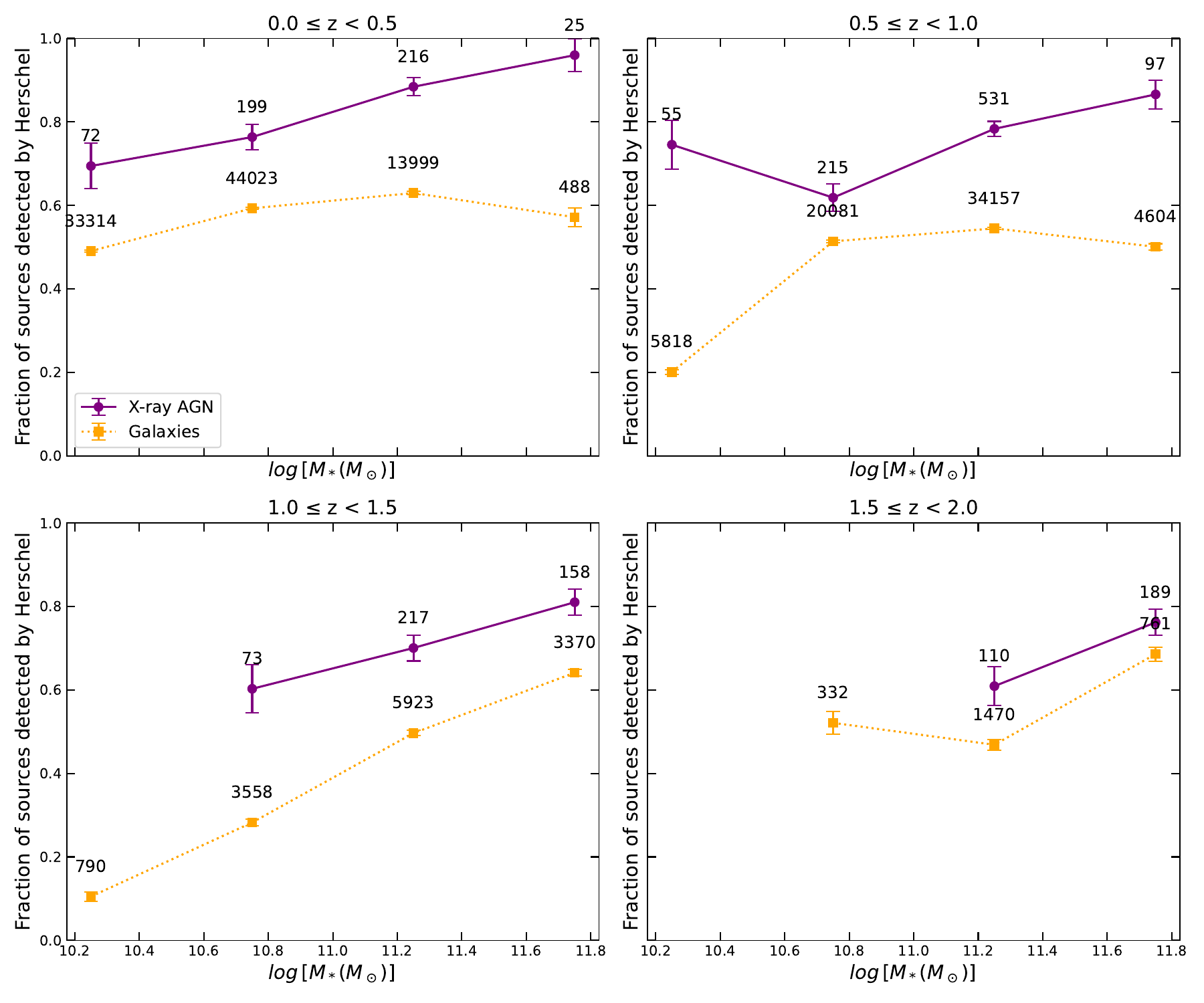} 
  \caption{Fraction of X-ray AGN (circles) and non-AGN galaxies (squares) detected by \textsl{Herschel} as a function of stellar mass. Error estimates are derived using bootstrap resampling. The total number of sources in each stellar mass bin is also displayed. Similar trends are observed in each of the five individual fields, although the magnitude of the AGN–non-AGN offset varies (see Appendix \ref{appendix_fields_separate}.)}
  \label{fig_frac_mstar}
\end{figure}  

\begin{figure}
\centering
 \includegraphics[width=1.\columnwidth, height=9cm]{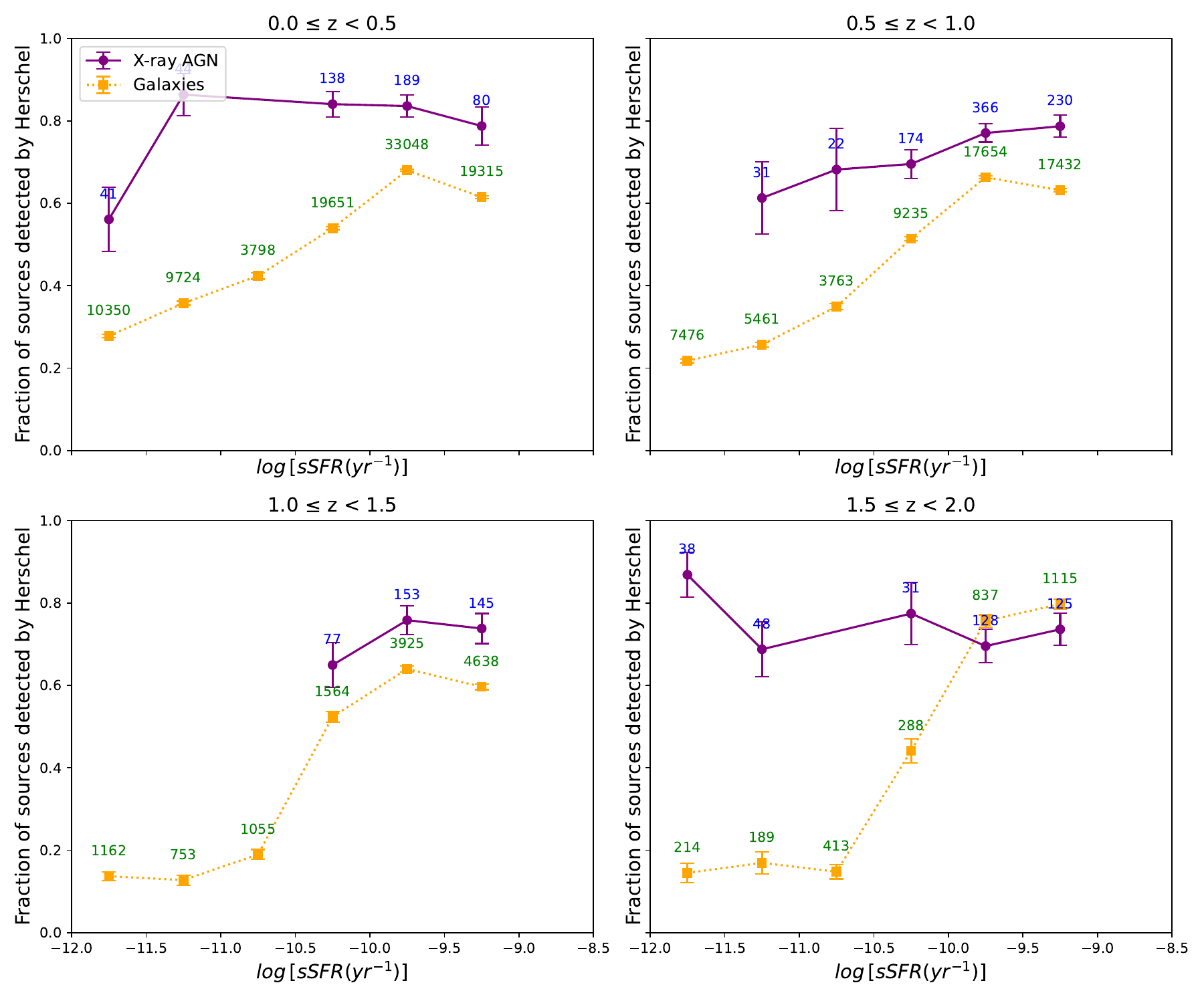} 
  \caption{Fraction of X-ray AGN (circles) and non-AGN galaxies (squares) detected by \textsl{Herschel} as a function of sSFR ($\rm sSFR=\frac{SFR}{M_*}$). Error estimates are derived using bootstrap resampling. The total number of sources in each stellar mass bin is also displayed. Similar trends are observed in each of the five individual fields, although the magnitude of the AGN–non-AGN offset varies (see Appendix \ref{appendix_fields_separate}.)}
 \label{fig_frac_ssfr}
\end{figure}  

\begin{figure}
\centering
  \includegraphics[width=0.75\columnwidth, height=4.5cm]{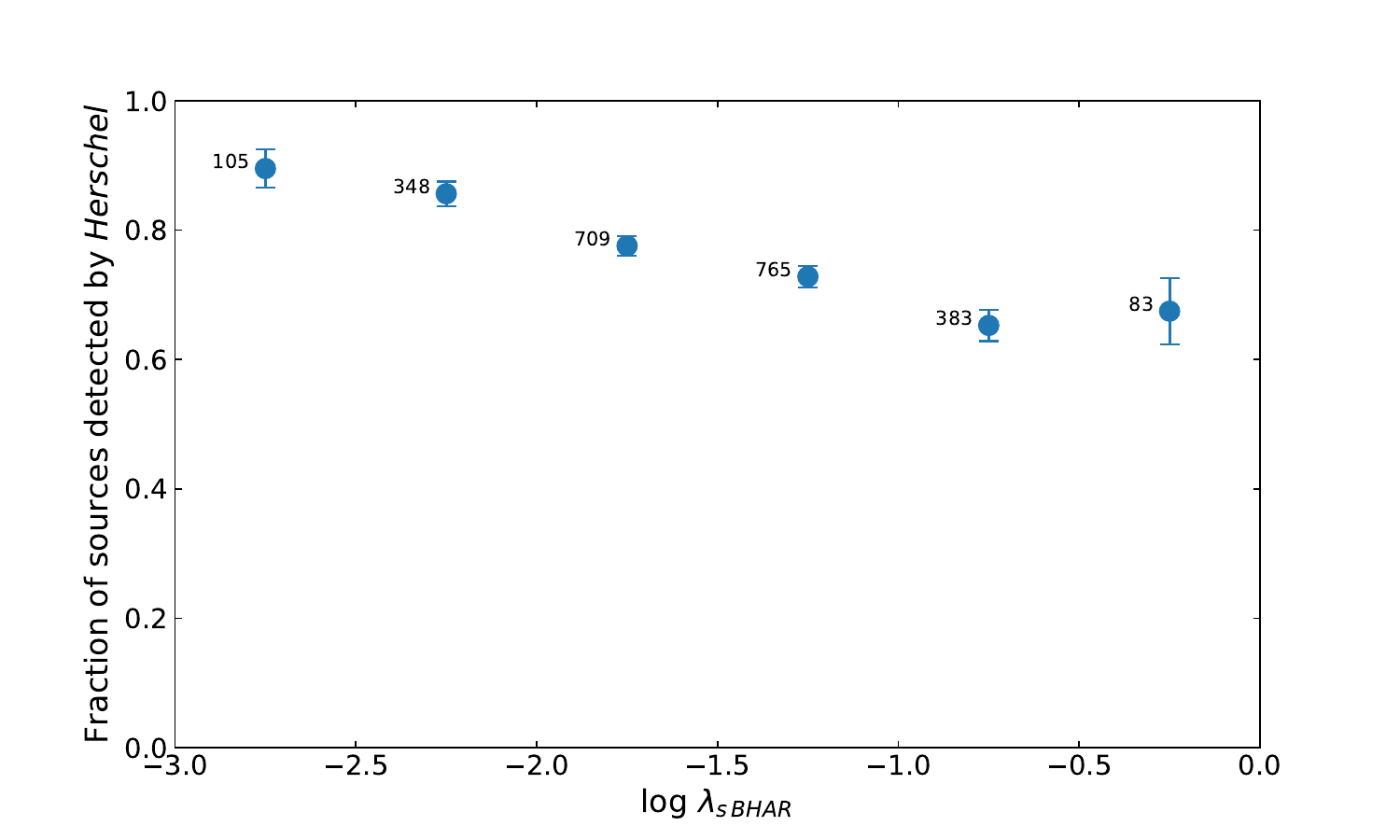} 
  \caption{Fraction of X-ray AGN detected by \textsl{Herschel} as a function of specific black hole accretion rate, $\lambda_{\text{sBHAR}}$. Error estimates are derived using bootstrap resampling. The total number of sources in each bin is also displayed. A similar trend is also observed in each of the five individual fields (see Appendix \ref{appendix_fields_separate}.)}
  \label{fig_frac_lambda}
\end{figure}  

\begin{figure}
\centering
  \includegraphics[width=0.8\columnwidth, height=5.2cm]{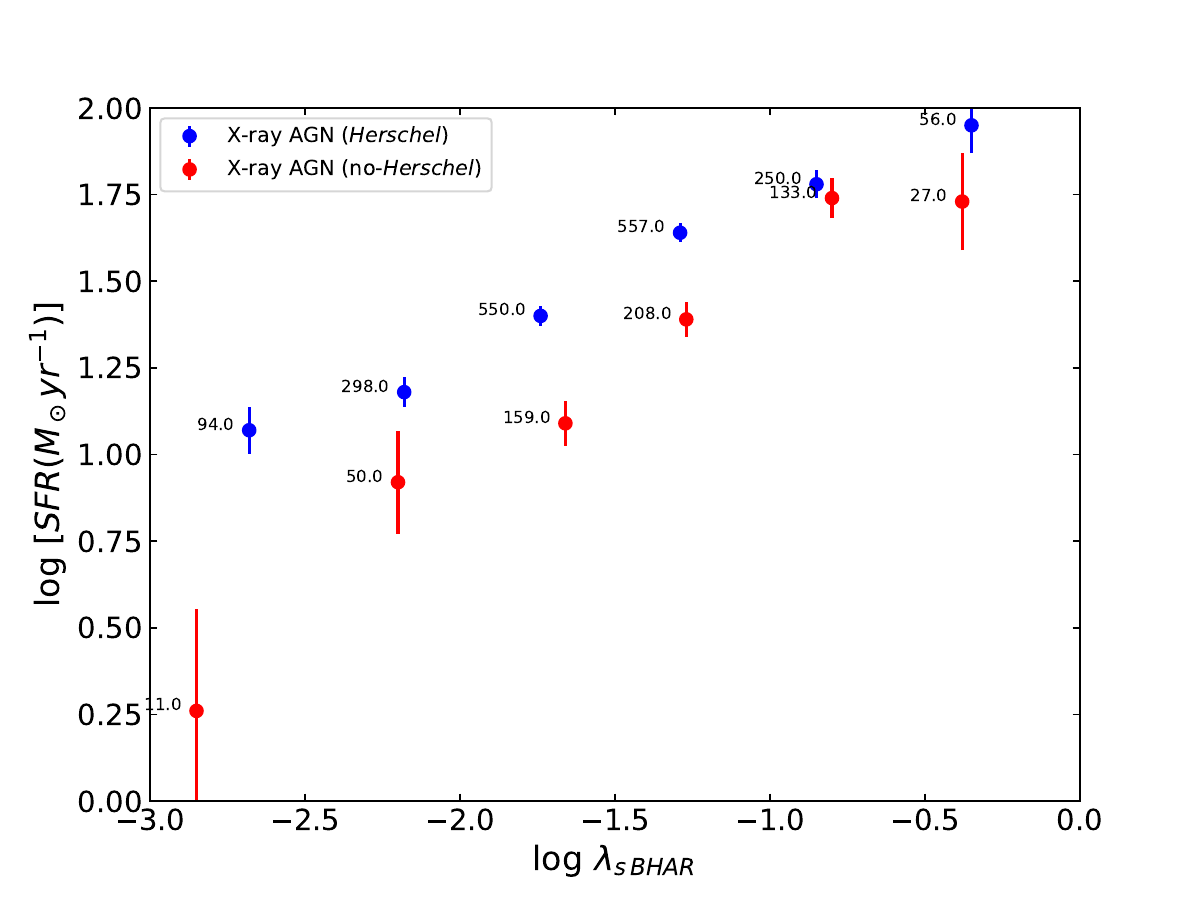} 
  \includegraphics[width=0.8\columnwidth, height=5.2cm]{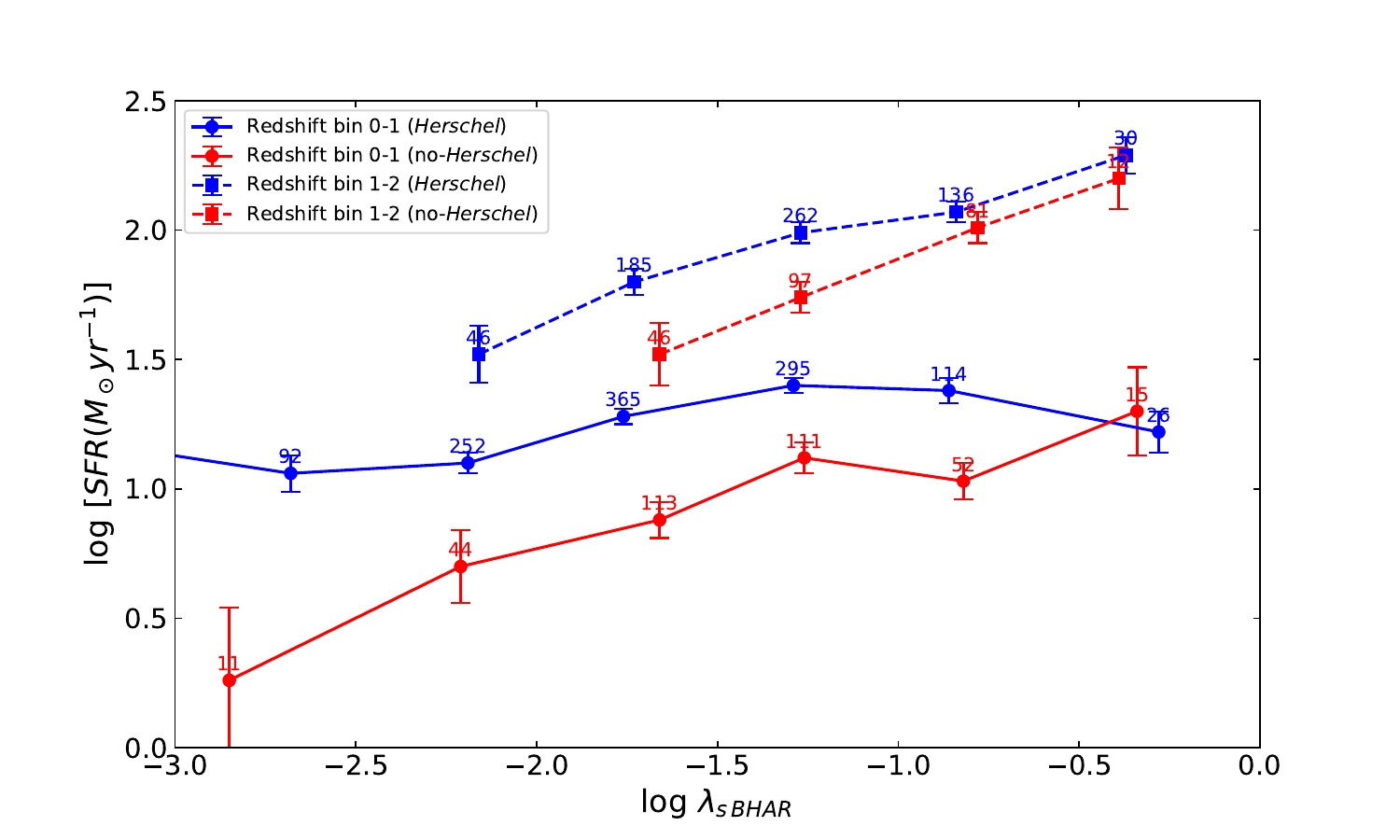} 
  \caption{SFR as a function of $\lambda_{\text{sBHAR}}$, for \textsl{Herschel} detected and no-\textsl{Herschel} detected X-ray AGN. The top panel presents the results in the total redshift range, $\rm 0<z<2$, while the bottom panel presents the measurements when we split the X-ray AGN sample into two redshift bins. }
  \label{fig_sfr_lambda}
\end{figure}  

\begin{figure*}[htbp]
\centering
  \includegraphics[width=1.2\columnwidth, height=7.5cm]{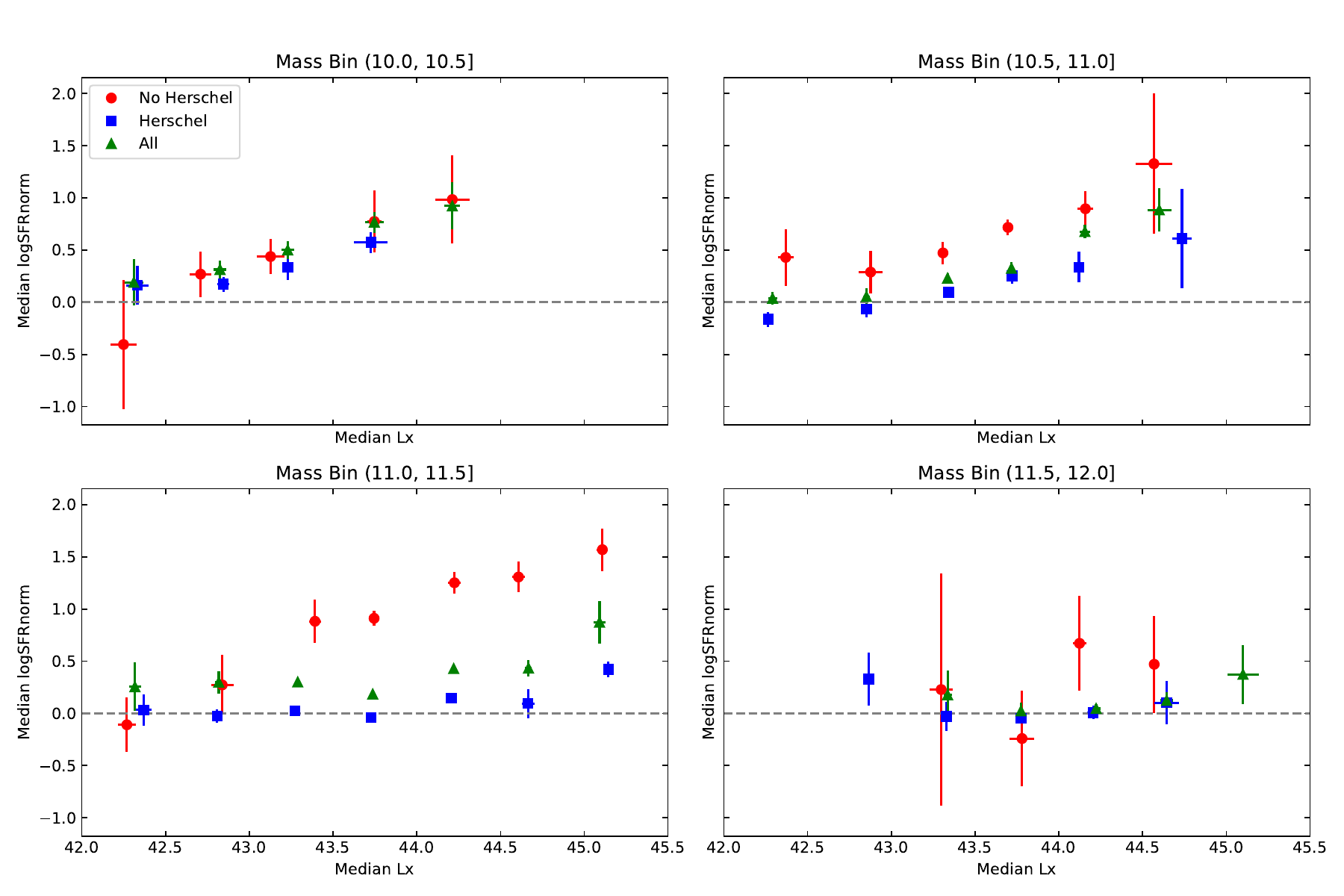}   
  \caption{SFR$_{norm}$ as a function of L$_X$, for four M$_*$ bins. Median values of SFR$_{norm}$ and L$_X$ shown. The bins are grouped by L$_X$ with a width of 0.5 dex, and the errors are $1\sigma$, calculated using bootstrap resampling. Only bins with ten or more sources are shown.}
  \label{fig_sfrnorm_lx_all}
\end{figure*}

\begin{figure}[htbp]
\centering
  \includegraphics[width=0.8\columnwidth, height=4.5cm]{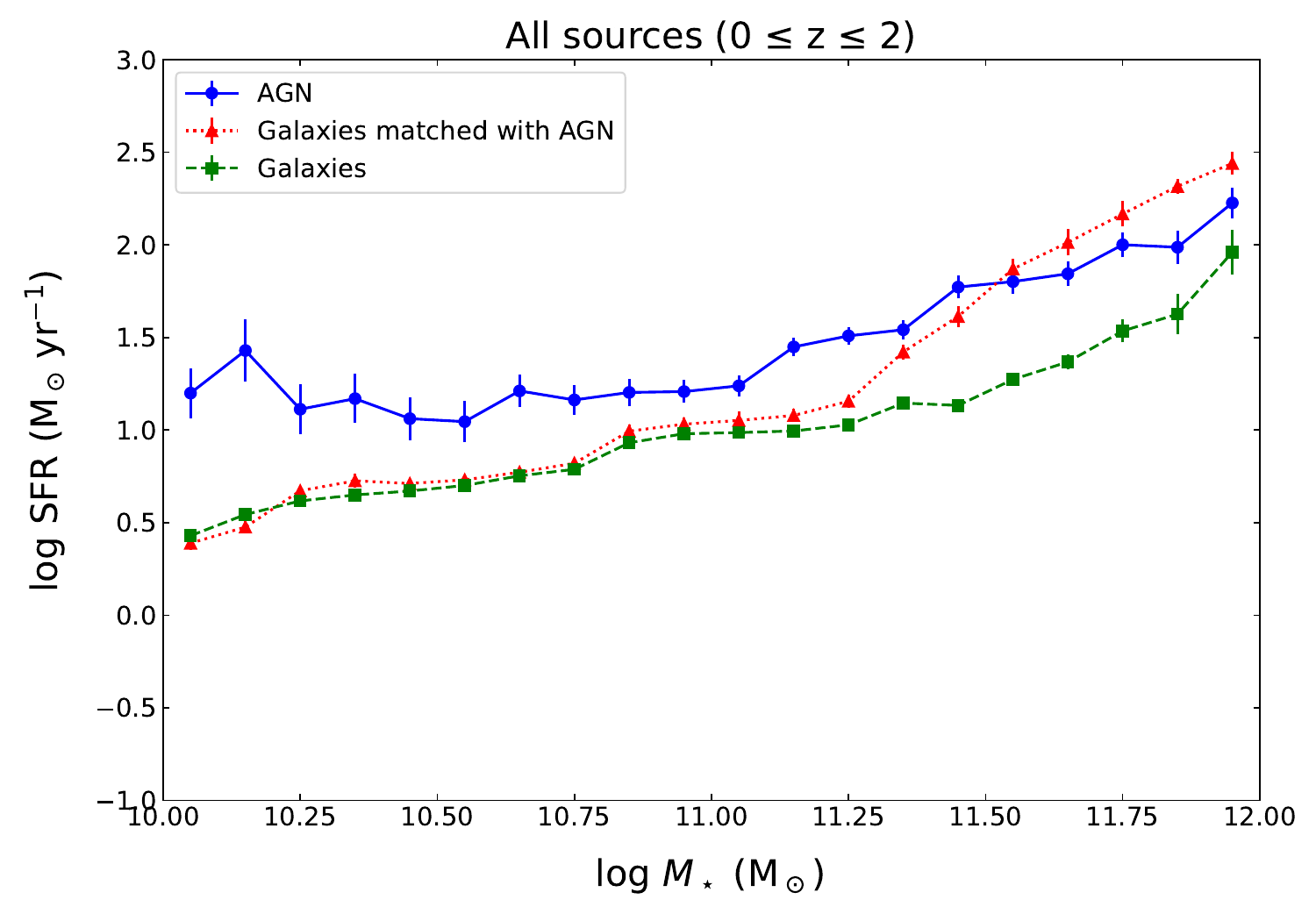}   
  \includegraphics[width=0.8\columnwidth, height=4.5cm]{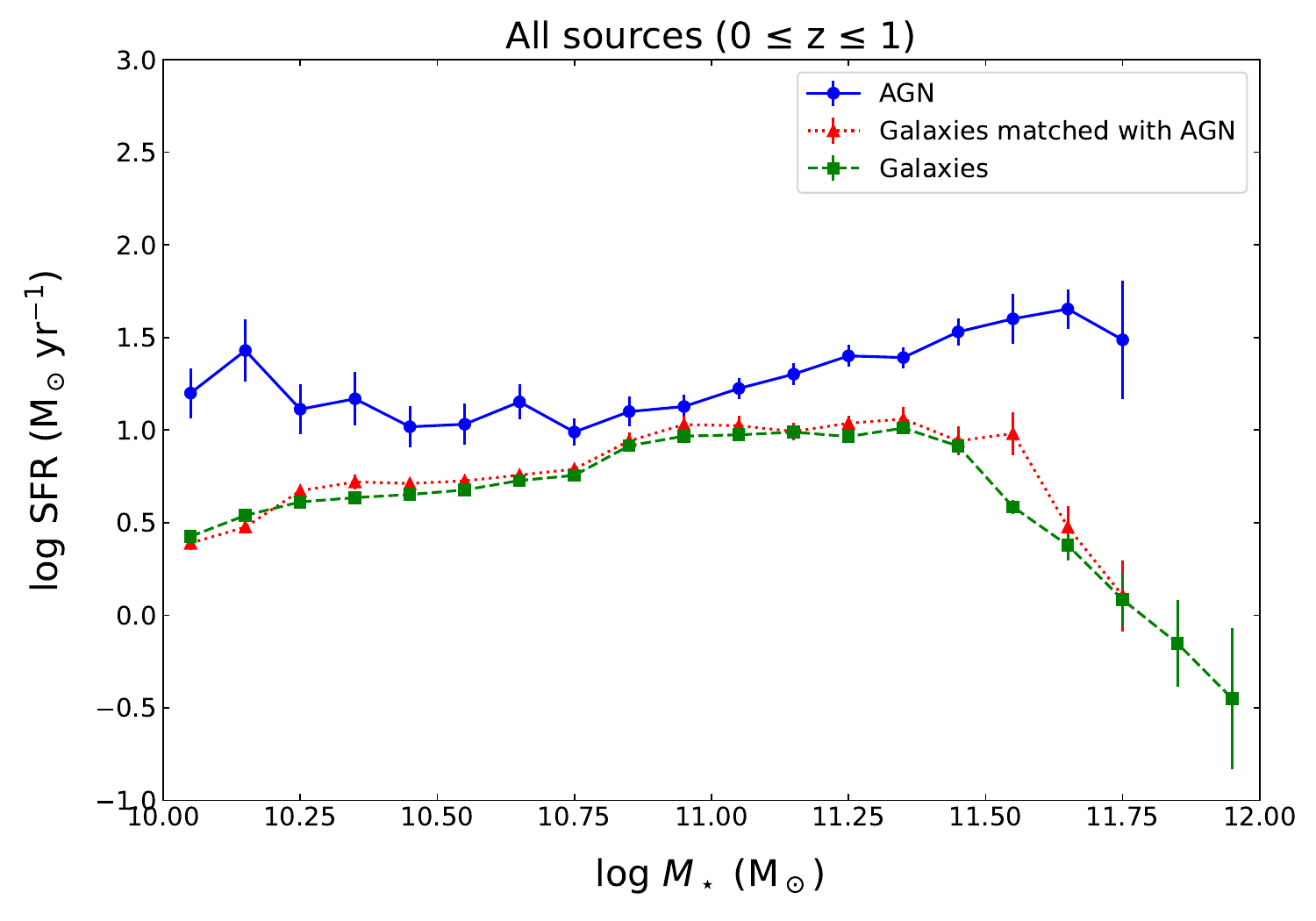} 
  \includegraphics[width=0.8\columnwidth, height=4.5cm]{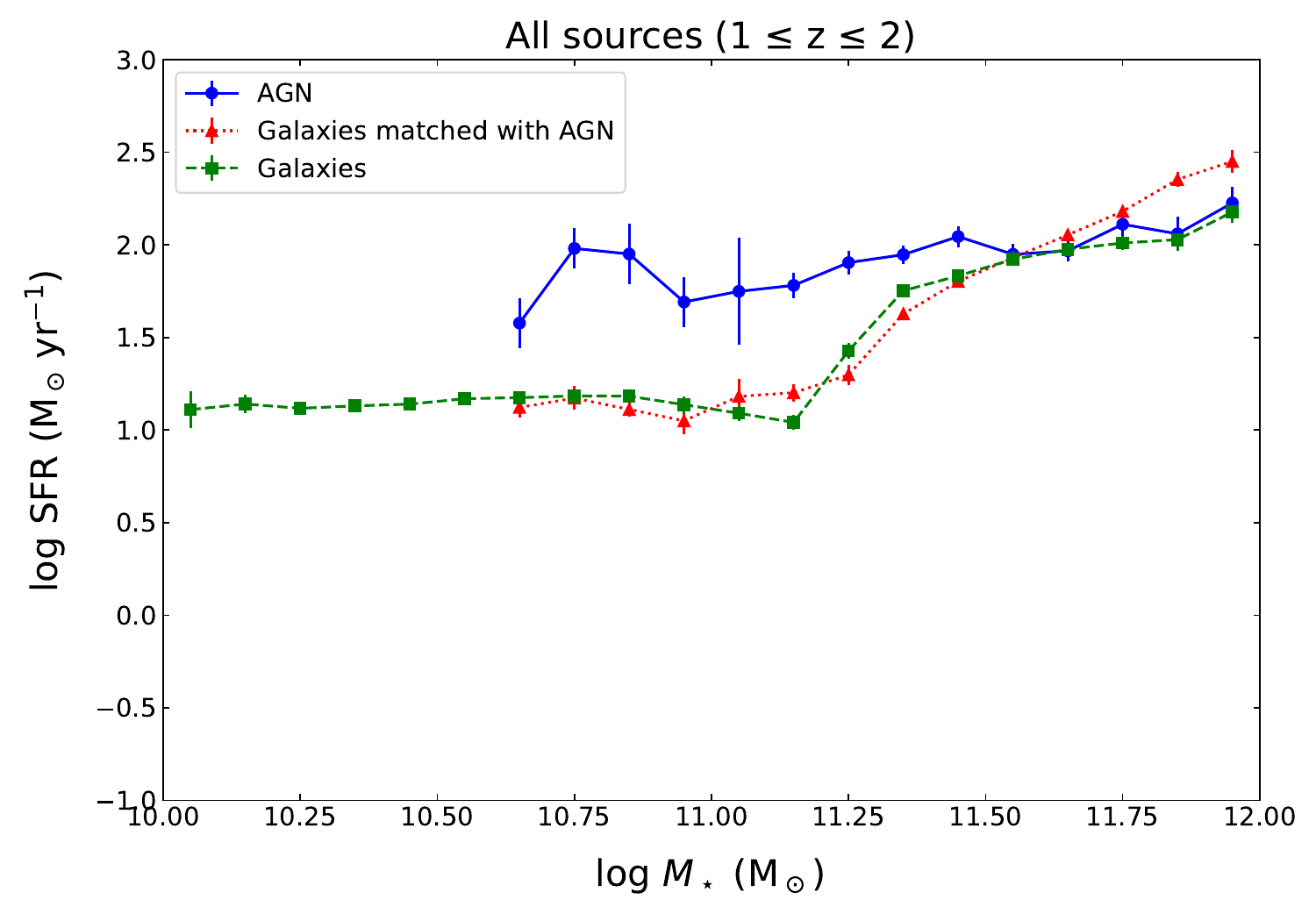} 
  \caption{SFR vs M$_*$ in logarithmic scales, for X-ray AGN, galaxies and AGN-matched galaxies (see text for more details). The top panel presents the results for the full redshift range utilized in our analysis. The middle and bottom panels, show the measurements at low and high redshift, respectively. At
$\rm z < 1$ (middle panel), the most massive non-AGN galaxies ($\log,[M(M_\odot)]>11.5$) experience a
pronounced SFR drop, about 1–2 dex lower than their AGN counterparts.}
  \label{fig_sfr_mstar}
\end{figure}

\subsection{Herschel detection criterion}
\label{sec_herschel_crit}

As mentioned previously, the primary objective of this study is to investigate the fraction of sources detected in the far-infrared spectrum (\textsl{Herschel}) as a function of stellar mass, Eddington ratio, and redshift and for different galaxy populations. These populations are defined based on whether they host an active SMBH, their optical classification (type 1 vs. type 2) and their X-ray absorption level.

We define a source as FIR-detected if it has a signal-to-noise ratio greater than 3 in at least one of the SPIRE bands (250, 350, or 500\,$\mu$m) of \textsl{Herschel}. Among the X-ray AGN, there are 2\,233 sources (approximately 73\%) detected by \textsl{Herschel}. In the non-AGN galaxy sample, there are 100,541 sources (approximately 44\%) detected by \textsl{Herschel}.

\subsection{$\rm K_s$ distributions}
\label{sec_ks_distrib}

Fig. \ref{fig_ks_4datasets} illustrates the distributions of the $\rm K_s$ magnitudes for \textsl{Herschel}-detected and non-detected sources (non-AGN galaxies and X-ray AGN). A secondary peak at fainter magnitudes ($\rm K_s > 22$) is observed among sources not detected by \textsl{Herschel}. This feature persists even when the sources are separated into X-ray AGN and non-X-ray galaxies. Upon further investigation, this secondary peak is found to be dominated by sources from the COSMOS field. These fainter sources exhibit higher SFR (by approximately 0.5\,dex), lower stellar masses (by approximately 0.3\,dex), and predominantly lie at higher redshifts ($\rm z > 1$, with a median value of 1.35) compared to their counterparts with $\rm K_s < 22$, both for X-ray AGN and non-AGN systems. To mitigate potential biases in our analysis, these sources are excluded from the datasets, by applying a $\rm K_s$ cut at 22. After this exclusion, the samples comprise 180,975 non-X-ray galaxies, of which 93,824 (approximately 52\%) are detected by \textsl{Herschel}, and 2,752 X-ray AGN, of which approximately 70\% are \textsl{Herschel} detected.

\subsection{Final datasets and their main properties}
\label{sec_final_samples}

A final constraint applied to our datasets is limiting the stellar mass range to \(10 < \log [M_*(M_\odot)] < 12\). At lower M$_*$ the number of X-ray AGN is significantly lower, making statistical analysis challenging. Additionally, a lower M$_*$ limit enters the dwarf galaxy population, where AGN activity may follow different accretion mechanisms \citep[e.g.,][]{ManzanoKing2020, Siudek2023a}. At higher M$_*$, the number of galaxies drops sharply, reducing the statistical reliability of comparisons. This restriction results in 2\,417 X-ray AGN (with 1\,822 detected by Herschel, approximately 75\%) and 172\,697 non-AGN galaxies (with 91\,537 detected by Herschel, approximately 53\%).

Figures \ref{fig_main_prop_agn} and \ref{fig_main_prop_gals} illustrate the main properties of the sources in the X-ray AGN and non-X-ray galaxy datasets, respectively, for both \textsl{Herschel}-detected and non-\textsl{Herschel}-detected sources. Table \ref{table_main_properties} provides the median values of the main properties for the different subsets. Comparing the subsets with and without \textsl{Herschel} detection, we observe similar M$_*$ and redshift distributions both in the case of X-ray AGN and in the case of non-AGN galaxies. However, for galaxies, those detected by \textsl{Herschel} exhibit significantly higher SFR, by approximately 0.75 dex, compared to \textsl{Herschel} non-detected sources. In contrast, this difference is much smaller for AGN (approximately 0.15 dex), resulting in similar star formation activity levels between the two AGN subsets. Additionally, the two AGN datasets display comparable SMBH power (L$_X$).

\section{Far-infrared incidence of galaxies and X-ray AGN detected by \textsl{Herschel}}
\label{sec_fraction_all}

In this section, we present the results of our analysis on the fraction of sources detected by \textsl{Herschel}. This includes the non-AGN galaxy and X-ray AGN subsets presented in the previous sections, analyzed as a function of stellar mass, SFR, Eddington ratio and redshift. Throughout our analysis, in this and all other sections, we only consider and present bins that contain at least 10 sources. 

\begin{table*}
\caption{Properties of X-ray AGN included in the $\log\lambda_{\text{sBHAR}}$ bins presented in Fig.~\ref{fig_frac_lambda}.}
\centering
\begin{tabular}{lrrrrr}
$\log\lambda_{\text{sBHAR}}$ & no. of sources & $\log\,[L_{X}~(\mathrm{erg\,s^{-1}})]$ & $\log\,[M_*(M_\odot)]$ & $\log\,[\mathrm{SFR}\, (M_\odot\, \mathrm{yr}^{-1})]$ & redshift \\
\hline
(-3.0, -2.5] & 78 (11)   & 42.49 (42.31) & 11.17 (11.07) & 1.07 (0.26) & 0.32 (0.36) \\
(-2.5, -2.0] & 245 (49)  & 43.06 (42.98) & 11.25 (11.22) & 1.22 (0.99) & 0.54 (0.56) \\
(-2.0, -1.5] & 482 (157) & 43.59 (43.53) & 11.30 (11.20) & 1.42 (1.09) & 0.77 (0.73) \\
(-1.5, -1.0] & 510 (207) & 43.98 (43.87) & 11.28 (11.18) & 1.65 (1.38) & 0.92 (0.97) \\
(-1.0, -0.5] & 235 (132) & 44.44 (44.34) & 11.24 (11.10) & 1.83 (1.75) & 1.08 (1.20) \\
(-0.5, 0.0]  & 53 (27)   & 44.54 (44.37) & 10.92 (10.78) & 1.95 (1.73) & 1.18 (0.86) \\
\end{tabular}
\tablefoot{Median values are shown for sources detected by \textsl{Herschel}, with values for non-detected sources shown in parentheses.}
\label{table_lambda_combined}
\end{table*}

\subsection{The dependence of far-infrared incidence on stellar mass and star formation rate}
\label{sec_fraction_mstar_sfr}

We first examine the fraction of sources detected in the far-IR as a function of M$_*$, in different redshift intervals (Fig. ~\ref{fig_frac_mstar}). Across all redshift bins, X-ray AGN exhibit a higher far-IR detection rate than non-AGN galaxies. On average, approximately 70\% of AGN and 50\% of non-AGN systems are detected by \textsl{Herschel}, with the detection fraction showing little variation with M$_*$ or redshift—except for a modest dip in the 1–1.5 redshift bin. Similar trends are observed in each of the five individual fields, although the magnitude of the AGN–non-AGN offset varies, being more pronounced in COSMOS and Stripe82, and less so in XMM-LSS (see Appendix \ref{appendix_fields_separate}). These differences likely reflect varying \textsl{Herschel} depths across and within fields \citep{Oliver2012, Shirley2021}, but they do not affect the overall trends.

Our findings indicate that X-ray AGN exhibit higher star formation activity than non-AGN galaxies at a fixed M$_*$, up to $\rm z=2$. To explore this further, Fig. \ref{fig_frac_ssfr} presents the fraction of sources detected by \textsl{Herschel} as a function of specific star formation rate, sSFR (sSFR = $\frac{SFR}{M_*}$) for AGN (circles) and galaxies (squares). We observe that the far-IR detection rate of X-ray AGN remains relatively constant, regardless of the host galaxy's star formation activity, across all redshifts probed. In contrast, for non-AGN galaxies, sources with higher star formation activity are more frequently detected by \textsl{Herschel}. These trends persist when the incidence of AGN and non-AGN galaxies is examined as a function of sSFR across each of the five individual fields (see Appendix \ref{appendix_fields_separate}).

The high far-IR detection fraction of X-ray AGN suggests that these systems reside in gas-rich environments capable of supporting both star formation and black hole growth. Since far-IR emission traces dust-obscured star formation, its prevalence among AGN hosts indicates that they retain substantial cold gas reservoirs. Notably, X-ray AGN remain detectable in the far-IR across all sSFR bins, in contrast to non-AGN galaxies, whose far-IR detectability strongly depends on sSFR. This pattern aligns with AGN feedback models and observational studies in which black hole accretion regulates, rather than abruptly quenches, star formation \citep[e.g.,][]{Harrison2017, Masoura2018, Scholtz2023}. While non-AGN galaxies show declining dust emission with lower sSFR, AGN hosts maintain high far-IR detection rates even at low sSFR—suggesting that AGN-driven processes may help retain or replenish cold gas. This supports a scenario in which AGN feedback operates over extended timescales, gradually modulating the cold gas supply before quenching ultimately takes effect.

Radio AGN studies have shown that low-excitation radio galaxies (LERGs) are found predominantly in massive, quiescent galaxies and are powered by hot gas accretion \citep[e.g.,][]{Hardcastle2007, Smolcic2009}. The incidence of LERGs in massive galaxies is roughly constant across cosmic time, suggesting that hot halo gas can sustain low-level AGN activity regardless of the star formation state of the host galaxy \citep[e.g.,][]{Best2012, Kondapally2025}. However, in lower-mass galaxies (log\, M${_*}$/M$_{\odot}$ < 11), LERG activity is significantly more enhanced in star-forming galaxies, suggesting that an additional cold-gas-driven fueling mechanism is required to sustain AGN activity in these systems \citep[e.g.,][]{Sabater2019}.

Interestingly, high-excitation radio galaxies (HERGs), which require higher accretion rates, are more similar to X-ray AGN in that they are likely cold-mode accretors and exhibit stronger correlations with the presence of cold gas and star formation. Given that X-ray AGN maintain a high far-IR incidence regardless of sSFR, our results suggest that X-ray AGN and HERGs may share similar fueling mechanisms, reliant on cold gas accretion. This further supports the hypothesis that cold inflows from the ISM and interactions play a dominant role in fueling X-ray AGN, rather than hot gas from the galaxy’s halo.

Our results may be consistent with an interpretation in which X-ray AGN represent an intermediate evolutionary phase between radiatively efficient AGN (e.g. HERGs) and radiatively inefficient AGN (e.g. LERGs). This scenario is supported by studies of HERG–LERG dichotomy in the local universe, where HERGs are typically hosted by bluer, star-forming galaxies, while LERGs reside in massive, quiescent galaxies and accrete at lower Eddington rates \citep[e.g.,][]{Best2012}. Further radio‑based diagnostics are required to explore this evolutionary connection in detail.

\subsection{The dependence of far-infrared incidence on Eddington ratio}
\label{sec_fraction_lambda}

In the case of X-ray AGN, we also examine the fraction of \textsl{Herschel} sources as a function of the Eddington ratio, n$_{\text{Edd}}$, which is defined as the ratio of the AGN bolometric luminosity, $L_{\text{bol}}$, to the Eddington luminosity, $L_{\text{Edd}}$ ($L_{\text{Edd}} = 1.26 \times 10^{38} \frac{M_{\text{BH}}}{M_\odot} \text{erg s}^{-1}$, where $M_{\text{BH}}$ is the mass of the SMBH). When $M_{\text{BH}}$ is not available, the specific black hole accretion rate, $\lambda_{\text{sBHAR}}$, is often used as a proxy for n$_{\text{Edd}}$ \citep[e.g.,][]{Aird2012, Georgakakis2017, Aird2018, Mountrichas2021c, Mountrichas2022a, Pouliasis2022}. The $\lambda_{\text{sBHAR}}$ is calculated using the following expression:

\begin{equation}
\lambda_{\text{sBHAR}} = \frac{k_{\text{bol}} L_{X,2-10\,\text{keV}}}{1.26 \times 10^{38}\, \text{erg s}^{-1} \times 0.002 \frac{M_*}{M_\odot}},   
\label{eqn_lambda}
\end{equation}
where $\rm k_{\text{bol}}$ is a bolometric correction factor that converts the X-ray luminosity to AGN bolometric luminosity. We adopted a value of $\rm k_{\text{bol}} = 25$ \citep[e.g.,][]{Georgakakis2017, Aird2018, Mountrichas2021c, Mountrichas2022a}. Previous studies have found differences between n$_{\text{Edd}}$ and $\lambda_{\text{sBHAR}}$ due to the scatter in the M$_{\text{BH}}$-M$_*$ relation and the method used for calculating the AGN bolometric luminosity \citep[e.g.,][]{Torbaniuk2021, Lopez2023, Mountrichas2023d}. Furthermore, different values for $\rm k_{\text{bol}}$ have also been used in different studies \citep[e.g.,][]{Yang2018b} as well as a luminosity dependent $\rm k_{\text{bol}}$ \citep[e.g.,][]{Lusso2012}. Nonetheless, our analysis focuses on the dependence of AGN incidence in the far-IR as a function of $\lambda_{\text{sBHAR}}$, rather than on the absolute values of $\lambda_{\text{sBHAR}}$ for the X-ray sources.

Fig. \ref{fig_frac_lambda} presents the detection rate of X-ray AGN in the far-infrared as a function of $\lambda_{\text{sBHAR}}$. We notice that the fraction of sources detected by \textsl{Herschel} is reduced with increasing $\lambda_{\text{sBHAR}}$.  A similar trend is observed when we split the X-ray dataset into two redshift bins, at $\rm z=1$. This is also the case across each of the five individual fields (see Appendix \ref{appendix_fields_separate}).

Table \ref{table_lambda_combined} presents the median values of L$_X$, M$_*$, SFR and redshift for the \textsl{Herschel} and non-\textsl{Herschel} sources, within the $\lambda_{\text{sBHAR}}$ bins shown in Fig. \ref{fig_frac_lambda}. We observe that, with the exception of the highest value bin, the increase in $\lambda_{\text{sBHAR}}$ across the bins is primarily driven by the increase in L$_X$, while M$_*$ remains relatively constant for both X-ray AGN populations. 

We also observe that, although the \textsl{Herschel} and \textsl{Herschel} non-detected X-ray AGN within the same $\lambda_{\text{sBHAR}}$ bins exhibit similar median values for L$_X$, M$_*$, and redshift, the SFR of sources detected in the far-IR is significantly higher than that of sources not detected in the far-IR. However, this difference diminishes at higher $\lambda_{\text{sBHAR}}$ values. This trend is illustrated in the top panel of Fig. \ref{fig_sfr_lambda}. A similar pattern is observed when the datasets are divided into two redshift bins, as shown in the bottom panel of Fig. \ref{fig_sfr_lambda}. This suggests that at high L$_X$ ($\log\,[L_{X,2-10keV}(\text{erg}\,\text{s}^{-1})] > 44$), both AGN populations (those detected by \textsl{Herschel} and those not detected) exhibit similar star formation activity. Furthermore, the SFR of both populations rises with increasing $\lambda_{\text{sBHAR}}$, in agreement with previous studies \citep[e.g.,][]{Torbaniuk2021, Torbaniuk2024}. We also notice that the SFR of X-ray AGN increases with L$_X$ while remaining consistent across similar M$_*$ values, regardless of their detection in the far-IR. \cite{Torbaniuk2024} reported a flattening of the $\lambda_{\text{sBHAR}}-$SFR relation, at high $\lambda_{\text{sBHAR}}$ values ($\rm log\,\lambda_{\text{sBHAR}}>0$), in the local universe ($\rm z<0.2$; see their Fig. 10). They suggest that this flattening could be caused by deviations in the stellar-to-BH mass scaling relation. Our X-ray dataset does not probe such high $\lambda_{\text{sBHAR}}$ values, but our results show a hint for a flattening in the $\lambda_{\text{sBHAR}}-$SFR relation at $\rm log\,\lambda_{\text{sBHAR}}>-1.0$, in particular at $\rm z<1$ (see Fig. \ref{fig_sfr_lambda}, bottom panel).

Our analysis reveals that the fraction of \textsl{Herschel} detected AGN decreases with increasing accretion efficiency ($\lambda_{\text{sBHAR}}$) and, consequently, with increasing L$_\mathrm{X}$ (Table~\ref{table_lambda_combined}). Conversely, the SFR of X-ray AGN host galaxies rises with $\lambda{\text{sBHAR}}$ (L$_\mathrm{X}$). These results suggest that high-L$_\mathrm{X}$ AGN remain in a phase where cold gas is still abundant, sustaining both supermassive black hole growth and star formation in the host galaxy. The moderate decline in far-IR detection with increasing $\lambda_{\text{sBHAR}}$ (L$_\mathrm{X}$) could indicate the onset of AGN feedback; however, rather than causing immediate quenching, it may signify a more gradual regulation of star formation. In this scenario, feedback from high-L$_\mathrm{X}$ AGN may redistribute gas rather than fully deplete it. AGN-driven turbulence or winds could compress gas in certain regions, triggering localized star formation while heating or expelling gas in others \citep[e.g.,][]{Cresci2015, Bieri2015, Maiolino2017, HermosaMunoz2024}. The feedback timescale may be long enough that these AGN continue to exhibit elevated SFR compared to non-AGN galaxies before quenching takes effect. The flattening of the $\lambda_{\text{sBHAR}}$–SFR relation observed in previous studies \citep[e.g.,][]{Torbaniuk2024} could represent an early indication that AGN feedback is beginning to regulate and limit future star formation.

The decline in Herschel detection rate with increasing accretion rate may appear counterintuitive in light of previous claims that powerful AGN significantly contribute to the far-IR or sub-mm emission. In particular, \citet{Symeonidis2016, Symeonidis2022} argued that dust heated by the AGN itself—rather than by star formation—could dominate the far-IR output in luminous QSOs, extending to wavelengths beyond 100\,$\mu$m. Such interpretations are often based on SFR calibrations using PAH features and mid-IR diagnostics. In contrast, our analysis employs full optical-to-far-IR SED fitting, allowing us to decompose AGN and star formation components and derive robust SFRs across a wide dynamic range. This approach minimizes contamination from AGN-heated dust and provides a more holistic view of the galaxy’s energy budget.

Interestingly, while we find that AGN are more likely to be detected in the far-IR than non-AGN galaxies of similar mass and redshift (Section~\ref{sec_fraction_all}), this enhanced detection does not scale with AGN power. Instead, our findings suggest that far-IR detection is primarily a tracer of cold gas content rather than AGN strength. AGN hosts may initially exhibit elevated dust and gas masses, facilitating both star formation and SMBH growth. However, as the accretion rate increases, feedback may begin to deplete or disrupt the cold gas reservoir, thereby reducing the far-IR emission even while SFR remains high. This could help explain the seemingly contradictory trends between detection rate and SFR, and potentially reconciles our results with models that predict either AGN suppression or contribution to the far-IR depending on the evolutionary phase. Our results are more consistent with studies such as \citet{Stanley2015} and \citet{Lani2017}, which find that the far-IR emission of AGN hosts is largely consistent with being star formation–dominated, rather than AGN-driven.

\section{The star formation of far-infrared detected and non-detected X-ray AGN and galaxies}
\label{sec_sfr}

To further investigate the star formation properties of X-ray AGN and non-AGN galaxies as a function of M$_*$ and their incidence in the far-IR, we employ the SFR$_\text{norm}$ parameter \citep[e.g.,][]{Mullaney2015, Masoura2018, Koutoulidis2022, Pouliasis2022}. SFR$_\text{norm}$ is calculated by dividing the SFR of each X-ray AGN by the SFR of non-AGN galaxies that closely match the AGN in terms of M$_*$ (within $\pm 0.2$ dex) and redshift (within $\pm 0.075 \times (1+z)$). When computing this average, each control galaxy is assigned a weight based on the inverse of the uncertainties in its SFR and M$_*$ estimates, as derived through the CIGALE methodology. The median of these weighted mean ratios is then adopted as SFR$_\mathrm{norm}$ for each M$_*$, L$_X$ bin, with associated errors estimated as 1\,$\sigma$, calculated via bootstrap resampling. Our measurements remain robust against the specific choice of matching region, although using smaller regions affects accuracy, as discussed in \cite{Mountrichas2021b}.

Fig. \ref{fig_sfrnorm_lx_all} presents SFR$_\text{norm}$ as a function of L$_X$ for different stellar mass bins. The total AGN population (green triangles) exhibits a nearly flat SFR$\text{norm}$ distribution at lower L$_X$, followed by an increase beyond a threshold that depends on M$_*$. This behavior is consistent with previous studies \citep[e.g.,][]{Mountrichas2021c, Mountrichas2022a, Mountrichas2022b, Mountrichas2024c, Cristello2024}. The absolute SFR$_\text{norm}$ values appear slightly higher than those in our prior works \citep[e.g.,][]{Mountrichas2022a, Mountrichas2022b, Mountrichas2024c}, as quiescent galaxies were not excluded from our sample. As noted in \cite{Mountrichas2021c}, this affects the normalization but not the underlying trends. Here, our primary goal is not to revisit the dependence of SFR$_\text{norm}$ on L$_X$ and M$_*$, but to distinguish between far-IR-detected and non-detected systems.

The results, shown in blue squares and red circles, correspond to \textsl{Herschel}-detected and non-detected sources, respectively. AGN consistently exhibit higher SFR than non-AGN galaxies in the non-\textsl{Herschel} detected sample across all M$_*$ and L$_X$ bins. However, among \textsl{Herschel}-detected systems, this trend holds only for lower-mass galaxies ($\log\,[M(M_\odot)]<11.0$), while at higher stellar masses, the SFR of AGN and non-AGN galaxies are indistinguishable.

To further explore these trends, we present the SFR–M$_*$ relation in Fig. \ref{fig_sfr_mstar}, which provides a complementary view, independent of AGN accretion power (L$_X$). In addition to X-ray AGN and non-AGN galaxies, we also include results for galaxies matched with X-ray AGN for the calculation of the SFR$_{\rm norm}$ parameter above, represented by triangles. The top panel shows that AGN host galaxies have higher SFR than non-AGN systems for $\log,[M(M_\odot)]<11.5$, but this difference disappears at higher stellar masses. The bottom panel ($\rm 1 < z <2$) shows that the AGN–galaxy SFR
contrast persists at high redshift, whereas the middle panel ($\rm 0 < z < 1$) reveals that, at
$\rm z < 1$, the most massive non-AGN galaxies ($\log,[M(M_\odot)]>11.5$) experience a
pronounced SFR drop, about 1–2 dex lower than their AGN counterparts.
 
This suggests that while AGN activity is often linked to enhanced star formation in lower-mass galaxies, it does not significantly impact the star formation of the most massive systems ($\log[M_*(M_\odot)]>11.5$). Furthermore, at low redshift ($\rm z<1$), the most massive non-AGN galaxies undergo accelerated quenching, whereas at higher redshifts ($\rm z>1$), they maintain high star formation activity. This could be attributed to the greater cosmic gas supply at earlier epochs, with a higher accretion rate of cold gas onto galaxies from the cosmic web. Even massive galaxies were able to sustain star formation, preventing the strong quenching observed at later times. However, at $\rm z<1$, a combination of factors—such as gas depletion, morphological stabilization, and environmental processes—likely leads to more efficient star formation suppression, particularly in the most massive non-AGN galaxies.

Further insights into the SFR–M$_*$ relation are obtained when separating the sample based on far-IR detection. Specifically, we find that AGN host galaxies maintain higher SFR than their non-AGN counterparts in the non-\textsl{Herschel} detected population, whereas the two populations exhibit comparable star formation activity among \textsl{Herschel}-detected systems. These trends are consistent across redshifts and reinforce the picture discussed above. The corresponding plots, which illustrate this behavior, are shown in Appendix~\ref{appendix_sfr_mstar} (Figs.~\ref{fig_sfr_mstar_herschel} and \ref{fig_sfr_mstar_noH}).

Previous studies have emphasized the importance of L$_X$ and M$_*$ when comparing the SFR of AGN and non-AGN galaxies \citep[e.g.,][]{Mountrichas2022a, Mountrichas2022b, Mountrichas2024c, Cristello2024}. Our results highlight an additional critical factor: far-IR detection status. While AGN exhibit similar SFR regardless of their \textsl{Herschel} detection, non-AGN galaxies show significantly lower SFR when they lack far-IR detections.

This distinction is particularly important when computing SFR$_\text{norm}$. In previous analyses, SFR$_\text{norm}$ was derived by matching AGN and non-AGN galaxies without differentiating between \textsl{Herschel}-detected and non-detected sources. While those studies demonstrated the robustness of SFR measurements from SED fitting even in the absence of \textsl{Herschel} photometry \citep[][]{Mountrichas2022a, Koutoulidis2022}, our current analysis reveals that far-IR detection significantly impacts SFR comparisons, especially for non-AGN galaxies.

Therefore, combining \textsl{Herschel}-detected and non-detected sources when comparing the SFR of AGN and non-AGN galaxies can introduce biases, particularly in SFR$_\text{norm}$ calculations. This distinction may help reconcile discrepancies between studies that exclusively used \textsl{Herschel} detected sources \citep[e.g.,][]{Stanley2015} and those that included both populations in their analysis, regarding the relationship between AGN activity and host galaxy star formation.

\begin{figure}
\centering
  \includegraphics[width=0.75\columnwidth, height=4.2cm]{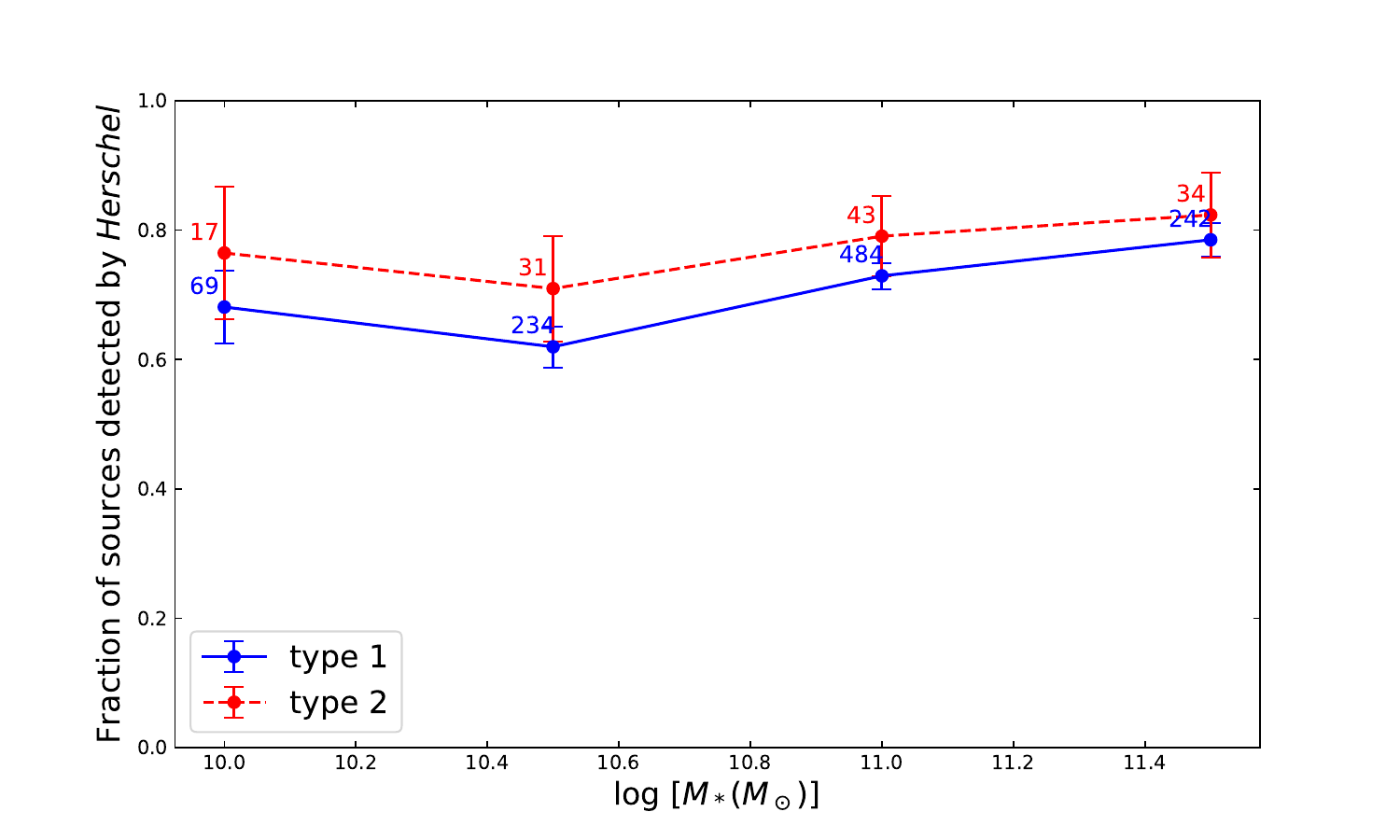} 
  \caption{Fraction of type 1 and 2 X-ray AGN detected by \textsl{Herschel} as a function of stellar mass. Measurements are grouped in M$_*$ bins of size 0.5\,dex. Error estimates are derived using bootstrap resampling. The total number of sources in each stellar mass bin is also displayed.}
  \label{fig_fraction_mstar_type}
\end{figure}

\begin{figure}
\centering
  \includegraphics[width=0.75\columnwidth, height=4.2cm]{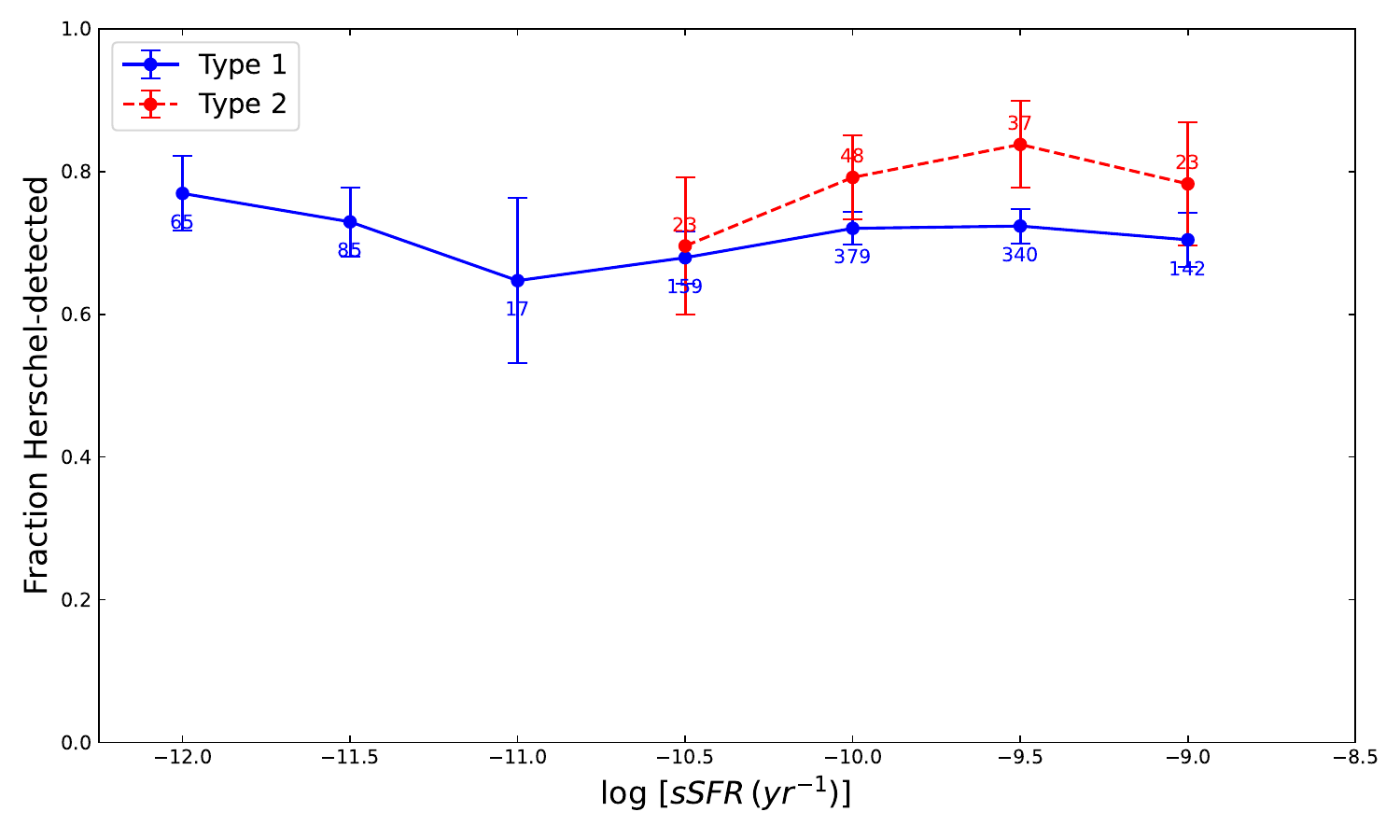} 
  \caption{Fraction of type 1 and 2 X-ray AGN detected by \textsl{Herschel} as a function of sSFR. Measurements are grouped in sSFR bins of size 0.5\,dex. Error estimates are derived using bootstrap resampling. The total number of sources in each bin is also displayed.}
  \label{fig_fraction_ssfr_type}
\end{figure}  

\begin{figure}
\centering
  \includegraphics[width=0.75\columnwidth, height=4.cm]{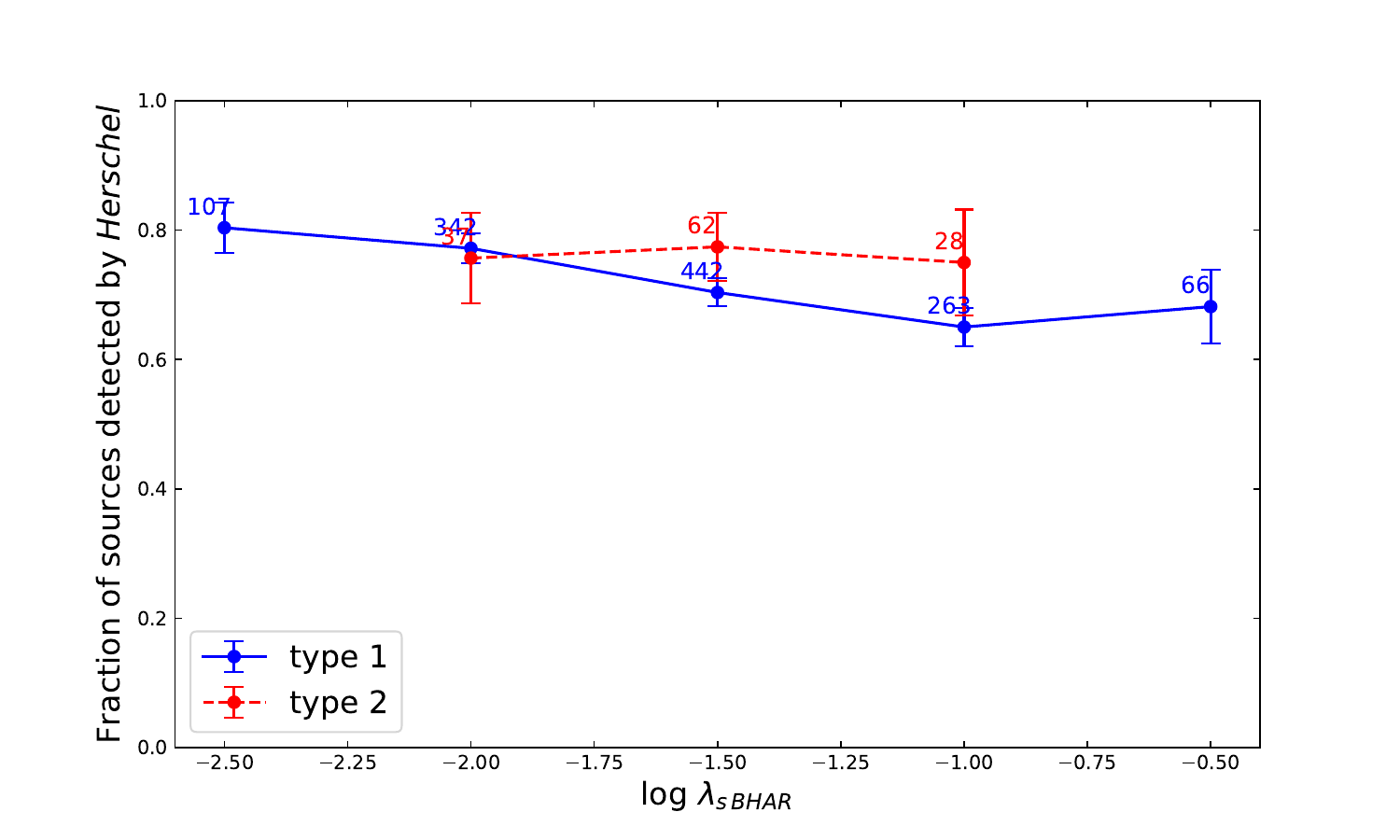} 
  \includegraphics[width=0.75\columnwidth, height=4.cm]{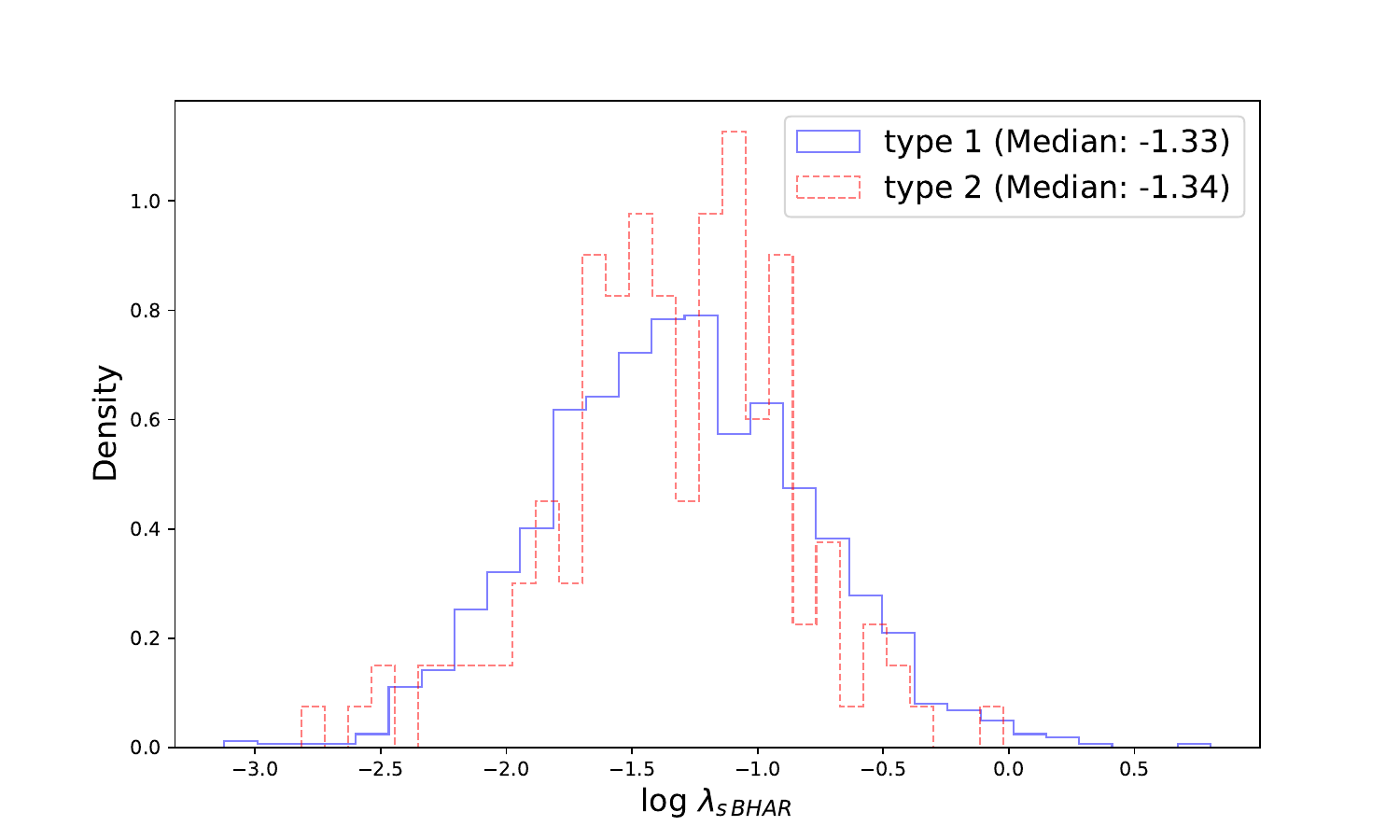} 
  \includegraphics[width=0.75\columnwidth, height=4.cm]{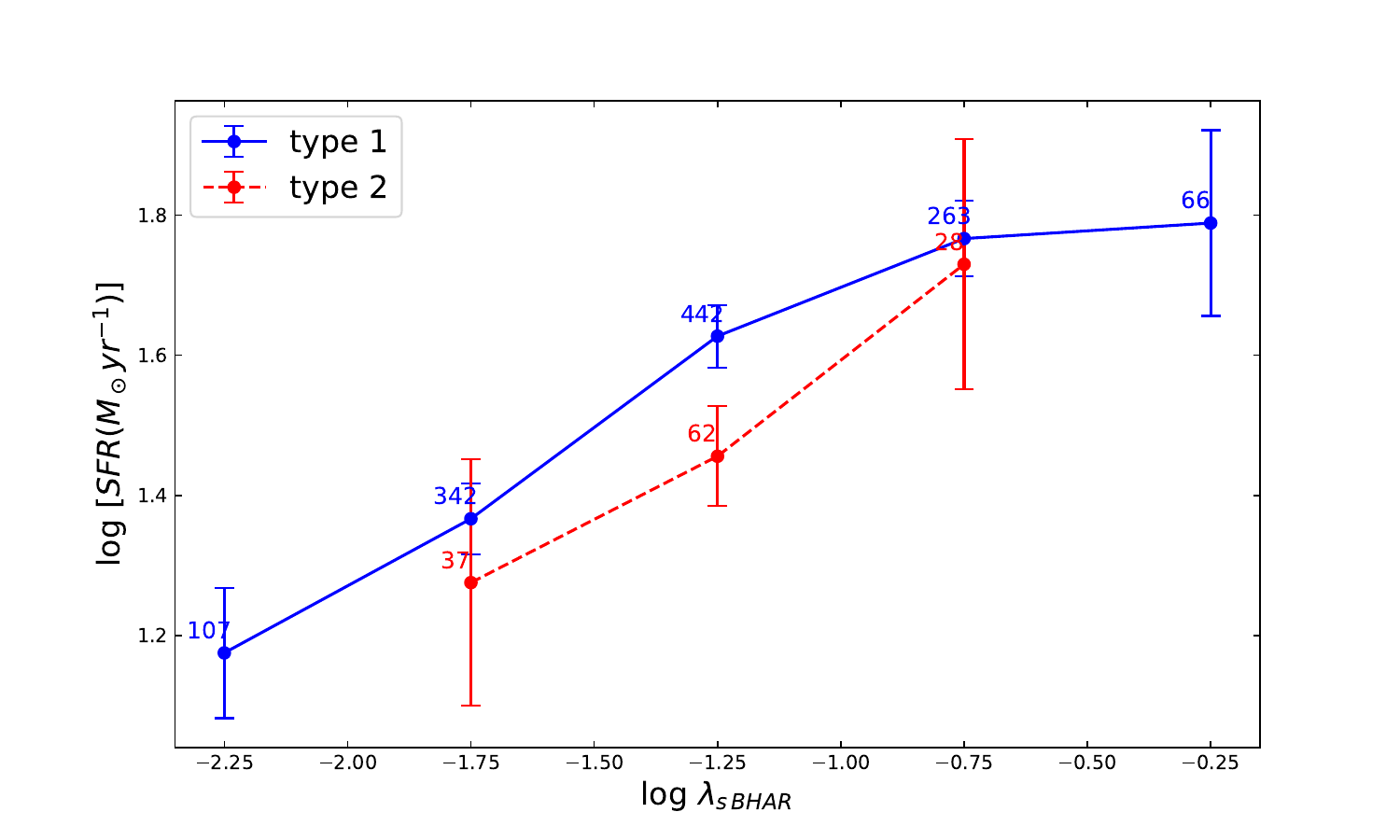} 
  \caption{
  The accretion efficiency, $\lambda_{\text{sBHAR}}$, of type 1 and 2 AGN. The top panel presents the fraction of type 1 and 2 X-ray AGN detected by \textsl{Herschel} as a function of $\lambda_{\text{sBHAR}}$. The middle panel shows the distribution of $\lambda_{\text{sBHAR}}$ for type 1 and 2 X-ray AGN. Application of KS-test yields a p$-$value of 0.39. The bottom panel illustrates the SFR of type 1 and type 2 X-ray AGN as a function of $\lambda_{\text{sBHAR}}$.}
  \label{fig_fraction_lambda_type}
\end{figure}

\section{The far-IR incidence of AGN categorized by obscuration and X-ray absorption}
\label{sec_agn_type}

In this section, we categorize AGN into obscured (type 2) and unobscured (type 1), as well as into X-ray absorbed and unabsorbed, and investigate the far-IR incidence of these AGN types as a function of M$_*$, SFR, $\lambda_{\text{sBHAR}}$, and redshift. Additionally, we analyze the levels of X-ray absorption in type 1 and type 2 AGN and explore the relationship between N$_H$ and accretion efficiency in these sources.

\subsection{AGN classification based on obscuration}
\label{sec_sed_classif}

We classify AGN into type 1 and type 2 based on SED fitting measurements, following the methodology outlined in our previous works \citep[][]{Mountrichas2021b, Mountrichas2024a, Mountrichas2024c}. The classification criteria are detailed in Appendix~\ref{appendix_agn_obscuration}. Applying these criteria yields 1,690 type 1 and 212 type 2 X-ray AGN.

\subsection{AGN classification based on X-ray absorption}

To examine the far-IR incidence of X-ray absorbed and unabsorbed AGN, we cross-matched our sample with the 4XMM AGN catalogue. Details of the cross-matching process, redshift validation, and absorption classification are provided in Appendix~\ref{appendix_agn_absorption}. This resulted in a subsample of 289 X-ray AGN, which we classified using a column density threshold of $N_{\mathrm{H}} = 10^{22}$\,cm$^{-2}$. Based on this criterion, 38 sources were identified as X-ray absorbed (87\% detected by \textsl{Herschel}) and 251 as unabsorbed (65\% \textsl{Herschel} detections).

\subsection{Results: Far-IR detection across AGN classifications}

Fig. \ref{fig_fraction_mstar_type} shows the detection rates in the far-IR for the type 1 and 2 AGN,  as a function of M$_*$. For both AGN classifications the fraction of \textsl{Herschel}-detected sources remains roughly constant with M$_*$. Type 2 AGN appear to have slightly higher far-IR detection rates by approximately 10\% consistently across all M$_*$ spanned by our data, however this difference is within the statistical uncertainties of the measurements. The slightly higher detection rate for type 2 AGN may suggest that these galaxies host more dust-obscured star formation, possibly linked to their higher dust and gas content compared to type 1 AGN. Similar trends are observed when we split the datasets into two redshift bins at $\rm z=1$.

In Fig. $\ref{fig_fraction_ssfr_type}$, we present the far-IR detection fraction of type 1 and type 2 AGN as a function of sSFR. For both AGN populations, the fraction of sources detected in the far-IR remains approximately constant with sSFR. However, type 2 AGN consistently exhibit higher far-IR detection rates than type 1 AGN at similar sSFR values. This trend, which mirrors the findings in Fig. $\ref{fig_fraction_mstar_type}$, suggests that type 2 AGN are more likely to reside in denser, dust-rich environments with obscured star formation, whereas type 1 AGN are found in less dust-enshrouded hosts with comparatively lower dust content.

The top panel of Fig. \ref{fig_fraction_lambda_type} illustrates the detection rates in the far-IR for the two AGN types as a function of $\lambda_{\text{sBHAR}}$. No statistically significant variations are observed in the detection fraction of both X-ray AGN populations in the far-IR as a function of $\lambda_{\text{sBHAR}}$. There is also an indication that type 2 X-ray AGN have higher detection fractions at the same $\lambda_{\text{sBHAR}}$ compared to type 1 AGN, although this difference is not statistically significant. Similar trends are found both at low ($\rm z<1$) and at high redshift ($\rm z>1$). Since sSFR and $\lambda_{\text{sBHAR}}$ both describe normalized growth rates (star formation and black hole accretion, respectively), their similar trends suggest a weak link between star formation and AGN activity in terms of detection fractions.

The middle panel of Fig. \ref{fig_fraction_lambda_type} presents the $\lambda_{\text{sBHAR}}$ distributions of both AGN types. The distributions appear similar, as indicated by the median values of $\lambda_{\text{sBHAR}}$, shown in the legend of the figure. A Kolmogorov-Smirnov (KS) test yields a p-value of 0.39, suggesting that the two distributions are not significantly different. Similar results are observed when the datasets are divided into two redshift bins, at $\rm z=1$.

In the bottom panel of Fig. \ref{fig_fraction_lambda_type}, we illustrate the SFR of type 1 and type 2 X-ray AGN as a function of $\lambda_{\text{sBHAR}}$. It is observed that type 1 AGN tend to exhibit higher star formation activity compared to type 2 AGN at similar $\lambda_{\text{sBHAR}}$. \cite{Mountrichas2024a} analyzed X-ray AGN in the eFEDS and COSMOS fields and found that type 1 X-ray AGN generally have higher SFR compared to type 2 AGN, particularly for sources at $\rm z<1$. At higher redshifts, their analysis indicated that the SFR of the two AGN types are similar, with type 2 AGN potentially having higher SFR at $\rm z>2$. Similar findings were reported by \cite{Mountrichas2024c} using X-ray AGN detected by the $\it{XMM-Newton}$ observatory.

The fact that type 2 AGN tend to have higher far-IR detection fractions but lower star formation rates at fixed $\lambda_{\text{sBHAR}}$ suggests that their enhanced far-IR emission is not solely a result of stronger star formation activity. Instead, the higher far-IR detection fraction indicates that type 2 AGN are more deeply embedded in dust-rich environments, where even moderate levels of star formation can produce significant far-IR emission. This suggests that obscuration in type 2 AGN is linked to the host galaxy environment rather than just the AGN itself. These findings challenge a purely orientation-based AGN unification model by indicating that large-scale galactic structures, such as mergers, high gas fractions, or turbulent interstellar media, play a significant role in shaping the observed AGN classifications and their infrared properties.

We note that similar trends are observed when we classify AGN based on their X-ray absorption. Specifically, for both AGN classifications the fraction of \textsl{Herschel}-detected sources remains roughly constant with M$_*$ and SFR. The \(\lambda_{\text{sBHAR}}\) distributions for X-ray absorbed and unabsorbed AGN are similar (KS-test p-value of 0.08). Previous studies have found that X-ray unabsorbed AGN tend to have higher \(\lambda_{\text{sBHAR}}\) values and the difference appeared to have high statistical significance. However, most of these studies used a higher \(N_H\) threshold to classify AGN and/or considered the uncertainties in \(N_H\) calculations \citep[e.g.,][]{Ricci2017a, Ananna2022, Georgantopoulos2023, Mountrichas2024b}. If we increase the \(N_H\) threshold to $N_H = 23\ \text{cm}^{-2}$, we are left with only four X-ray absorbed AGN.

\subsection{Obscuration vs. absorption}
\label{sec_obsc_vs_absorp}

While it is not the primary objective of this study, the available dataset provides an opportunity to examine the levels of X-ray absorption in type 1 and type 2 AGN. Accordingly, this section will explore the relationship between N$_H$ and accretion efficiency in these sources.

\begin{figure}
\centering
  \includegraphics[width=0.75\columnwidth, height=4.cm]{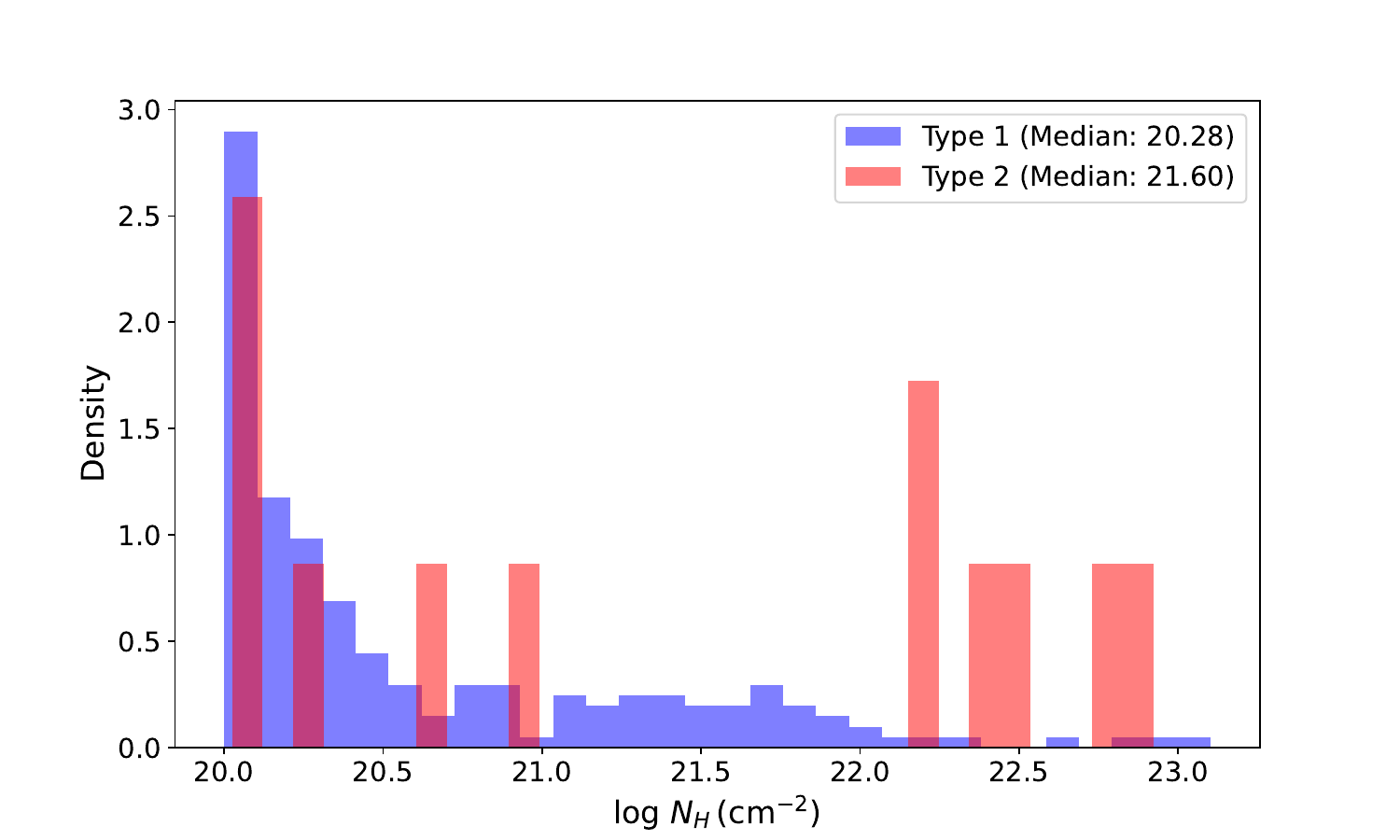} 
  \caption{Distribution of N$_H$ for type 1 and type 2 AGN.}
  \label{fig_nh_type12}
\end{figure}

Among the 289 X-ray AGN with available X-ray spectral measurements, 209 have a secure classification based on the SED fitting analysis (Sect. \ref{sec_sed_classif}). Specifically, 197 are classified as type 1 and 12 as type 2. Fig. \ref{fig_nh_type12} displays the distribution of N$_H$ for these two AGN types. The majority of type 1 AGN (188 out of 197, approximately 95\%) exhibit low N$_H$ values ($<22$), with a median value of $N_H = 20.3\ \text{cm}^{-2}$ . Conversely, six out of the 12 type 2 AGN have high N$_H$ values ($>22$) and a median value of $N_H = 21.6\ \text{cm}^{-2}$.

\begin{figure}
\centering 
  \includegraphics[width=0.95\columnwidth, height=6.3cm]{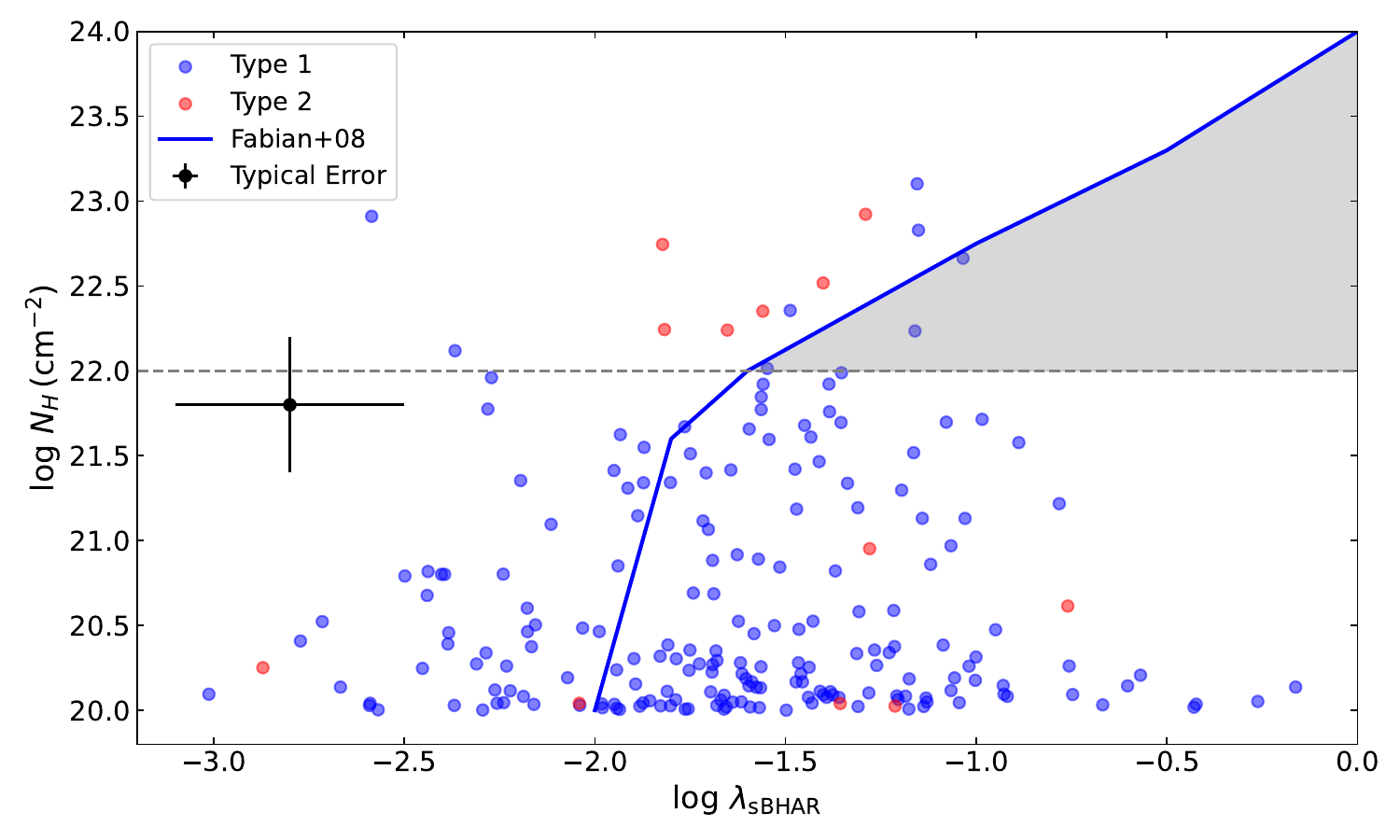} 
  \caption{Distribution of the type 1 and 2 AGN in the \( N_\mathrm{H} - \lambda_{\text{sBHAR}} \) space. The horizontal dashed line at \(\log N_\mathrm{H} = 22\,\mathrm{cm}^{-2}\) marks the region where dust lanes from the host galaxy are typically expected to contribute to the line-of-sight column density. The blue curve represents the effective Eddington ratio, assuming a dust abundance of 0.3 times the galactic value \citep{fabian2008}. The blowout phase, i.e., the region in the \( N_\mathrm{H} - \lambda_{\text{sBHAR}} \) space where radiation pressure pushes away the obscuring material, is shown by the shaded area. Typical uncertainties are $\sim 0.4$\,dex in N$_H$ and $\sim 0.3$\,dex in $\lambda_{\text{sBHAR}}$ (in logarithmic scale in both cases). These should be taken into account when interpreting whether individual points fall above or below the threshold.}
  \label{fig_nh_lambda}
\end{figure}

Fig. \ref{fig_nh_lambda} illustrates the distribution of type 1 and type 2 AGN in the N$_H$-$\lambda_{\text{sBHAR}}$ space. The horizontal dashed line at $\log\,N_H = 22\, \text{cm}^{-2}$ demarcates the region where dust lanes from the host galaxy typically contribute to the line-of-sight column density. The blue curve represents the effective Eddington ratio, adopted from \cite{fabian2008}, assuming a dust abundance of 0.3 the galactic value \citep[e.g.,][]{Vijarnwannaluk2024}. The shaded area indicates the blowout phase, where radiation pressure is expected to expel the obscuring material \citep[e.g.,][]{Ricci2017}.

None of the type 2 X-ray AGN fall within the shaded area. This could suggest that the obscuration in our sample of type 2 AGN is likely due to a radiation-pressure regulated dusty torus \citep{Vijarnwannaluk2024}, similar to AGN in the local universe \citep{Ananna2022, Ricci2022}. We note, however, that the number of Type 2 AGN in this diagram is very small, and as such the observed placement near the proposed blow-out region may not be statistically robust. While this positioning is suggestive, more data are needed to confirm whether these sources trace a distinct evolutionary phase. As for type 1 AGN, as noted earlier, a small subset (nine sources) displays high N$_H$ values, with only three of these falling within the shaded area. This suggests that the radiative feedback from the AGN is highly effective at clearing out the immediate environment of the accreting black hole. High X-ray absorption in type 1 AGN has also been reported in previous studies \citep[e.g.,][]{Merloni2014, Liu2018, Masoura2020} and can be explained by scenarios involving dust-free gas \citep[e.g.,][]{Gallagher2006, Ichikawa2019, Kudoh2023}. 

It is important to note that, as previously mentioned, earlier studies have observed an offset between n$_{Edd}$ and its commonly used proxy, $\lambda_{\text{sBHAR}}$. For example, \cite{Lopez2023} reported that n$_{Edd}$ tends to be lower by 0.6 dex compared to $\lambda_{\text{sBHAR}}$, while \cite{Mountrichas2023d} identified a difference of 0.25 dex. In our analysis, using bolometric AGN luminosity calculations from SED fitting, rather than the bolometric correction $\rm k_{bol}\times L_X$, we find that $\lambda_{\text{sBHAR}}$  values are, on average, lower by 0.33 dex. Adjusting our $\lambda_{\text{sBHAR}}$  measurements presented in Fig. \ref{fig_nh_lambda} by this 0.33 dex would result in the removal of our sources from the shaded area.

\section{Conclusions}
\label{sec_conclusions}

The primary aim of this work study is to study the properties of X-ray AGN and non-X-ray galaxies, which do not host an active SMBH, as a function of their incidence in the far-infrared. We use data from five fields: COSMOS, XMM-LSS, Stripe82, ELAIS-S1, and CDFS-SWIRE. By cross-matching the galaxy datasets with the 4XMM-DR11s X-ray catalog and applying a luminosity threshold of \(\log\,[L_{X,2-10keV}(\text{erg s}^{-1})] > 42\), we identify galaxies hosting X-ray detected AGN. We construct and fit the SEDs for all sources, utilizing photometry across the optical, near-infrared, mid-infrared, and far-infrared wavelengths, with the latter sourced from \textsl{Herschel} observations. After applying several criteria, such as photometric coverage and the reliability of SED fitting measurements, we derived a final sample comprising 172,697 non-AGN galaxies (without evidence for an AGN either from X-ray detection or SED fitting), of which 91,537 (approximately 53\%) are detected by \textsl{Herschel}, and 2\,417 X-ray AGN, of which 1\,822 (approximately 75\%) are detected by \textsl{Herschel}. These datasets are mass-complete, with stellar masses ranging between \(\rm 10<log\,[M_*(M_\odot)]<12\) and redshifts within \(\rm 0<z<2\).

We first examined the detection rates of non-AGN galaxies and X-ray AGN in the far-infrared as a function of M$_*$, SFR and accretion efficiency and compared the SFR of the two populations as a function of their incidence in the far-IR. Our key conclusions are as follows:

\begin{itemize} 

\item[$\bullet$] The incidence of X-ray AGN in the far-IR is higher than that of non-AGN galaxies across all M$_*$ and sSFR covered by our dataset (Figures \ref{fig_frac_mstar} and \ref{fig_frac_ssfr}).

\item[$\bullet$] The fraction of both populations detected in the far-IR remains relatively stable across M$_*$. A similar trend is observed for X-ray AGN as a function of sSFR. In contrast, among non-AGN galaxies, sources with higher star formation activity are more frequently detected by Herschel (Figures \ref{fig_frac_mstar} and \ref{fig_frac_ssfr}).

\item[$\bullet$] The fraction of X-ray AGN detected by \textsl{Herschel} is reduced with increasing accretion efficiency (Fig. $\ref{fig_frac_lambda}$).

\item[$\bullet$] The SFR of both \textsl{Herschel} detected and \textsl{Herschel} non-detected X-ray AGN increases with the specific black hole accretion rate (Fig. \ref{fig_sfr_lambda}).

\item[$\bullet$] Among galaxies not detected by \textsl{Herschel}, X-ray AGN exhibit significantly higher SFR compared to non-AGN galaxies, across the entire X-ray luminosity range probed in our study (\(\rm 42 < \log\,[L_{X,2-10keV}(erg\,s^{-1})] < 45\)).

\item[$\bullet$] For \textsl{Herschel}-detected sources, X-ray AGN display enhanced SFR relative to non-AGN galaxies only at low stellar masses (\(\rm \log\,[M_*(M_\odot)] < 10.5\); Fig.~\ref{fig_sfrnorm_lx_all}). This likely reflects the fact that in X-ray AGN, the SFR increases with stellar mass regardless of far-IR detection, whereas in non-AGN galaxies, the SFR–M$_*$ relation is positive only for \textsl{Herschel}-detected sources and remains flat for far-IR undetected ones, at least up to \(\rm log\,[M_*(M_\odot)]<11.5\) (Figures \ref{fig_sfr_mstar}, \ref{fig_sfr_mstar_herschel}, \ref{fig_sfr_mstar_noH}).

\end{itemize} 

One possible interpretation of our results is that X-ray AGN reside in cold gas-rich environments where both star formation and black hole accretion are fueled by gas inflows. The sustained far-IR detection of X-ray AGN across all sSFR bins suggests that AGN activity can coexist with ongoing star formation, potentially regulating it rather than abruptly quenching it. This would be consistent with models in which AGN feedback modulates the gas supply over time. The observed trends are also reminiscent of those reported for HERGs in the radio regime, which are thought to be fueled by cold gas, in contrast to LERGs that rely on hot gas accretion. Although our study does not include direct radio diagnostics, future work incorporating radio-based AGN classifications could test whether X-ray AGN indeed represent an intermediate evolutionary phase transitioning from cold to hot gas accretion as their host galaxies deplete their cold gas reservoirs.

Our analysis underscores the need to consider photometric coverage when comparing the star formation rates (SFRs) of X-ray AGN and non-AGN galaxies. While SED fitting tools like \textsc{CIGALE} can reliably estimate SFRs even without far-IR data \citep[e.g.,][]{Mountrichas2022c}, we find that AGN show similar SFRs regardless of far-IR detection. In contrast, non-AGN galaxies without far-IR coverage tend to have significantly lower SFRs. This difference suggests that far-IR non-detections in AGN are not necessarily linked to suppressed star formation, but may instead reflect variations in dust content or geometry, potentially influenced by AGN activity. As a result, far-IR detection in AGN may not directly trace SFR in the same way it does for galaxies, and this should be taken into account in comparative studies.

We, then, utilized SED fitting measurements to classify X-ray AGN into type 1 and type 2 categories, identifying 1\,690 type 1 and 212 type 2 AGN in our dataset. To explore the role of X-ray absorption, we analyzed 289 X-ray AGN with available \(N_H\) calculations derived from X-ray spectral fitting. Using a threshold of \(N_H = 10^{22}\ \text{cm}^{-2}\), these AGN were further categorized into X-ray absorbed (high \(N_H\)) and X-ray unabsorbed (low \(N_H\)) groups, with 38 classified as X-ray absorbed and 251 as X-ray unabsorbed.

Our analysis shows that both AGN types, regardless of classification method, exhibit similar trends in their far-IR incidence as a function of M$_*$, SFR, and $\lambda{\text{sBHAR}}$, aligning with the total AGN population. Obscured and unobscured X-ray AGN maintain a roughly constant far-IR detection rate across M$_*$, sSFR, and accretion efficiency, though obscured AGN display a slightly higher detection fraction (about 10\%). However, type 1 X-ray AGN tend to exhibit higher star formation activity than type 2 AGN at similar $\lambda_{\text{sBHAR}}$ (Figs. \ref{fig_fraction_mstar_type}, \ref{fig_fraction_ssfr_type}, \ref{fig_fraction_lambda_type}). These findings suggest that the enhanced far-IR emission in type 2 AGN is not solely due to stronger star formation but rather indicates that they are embedded in dust-rich environments where even moderate star formation can contribute significantly to far-IR emission. This supports the idea that obscuration in type 2 AGN is influenced by large-scale host galaxy conditions rather than just the AGN itself.

Finally, our findings suggest that the obscuration observed in type 2 sources likely arises from a combination of large-scale host galaxy dust and a radiation-pressure-regulated dusty torus, consistent with previous studies in the local universe. We note, however, that our type 2 AGN sample is small, containing only 12 sources, of which just 6 exhibit high levels of X-ray absorption (\( \log\,N_{\rm H} > 22 \)). For type 1 AGN, only a small fraction (nine out of 197) show high X-ray absorption, suggesting that in these cases the obscuration may arise from torus-independent material such as dust-free or low-dust gas (Figs. 13 and 14).

\section*{Data availability}

The data used in this study are based on publicly available surveys cited in Section 2. Processed data products and analysis scripts are available from the corresponding author upon reasonable request.

\begin{acknowledgements}
This project has received funding from the European Union's Horizon 2020 research and innovation program under grant agreement no. 101004168, the XMM2ATHENA project. FJC and SM acknowledge funding from grant PID2021-122955OB-C41 funded by
MCIN/AEI/10.13039/ 501100011033 and by ERDF A way of making Europe.
\end{acknowledgements}

\bibliography{mybib}
\bibliographystyle{aa}

\appendix

\section{SED fitting: modules, parameter grid, quality criteria, and identification of non-X-ray AGN}

Here we  present the templates and parameter grid used in the SED fitting, along with the quality criteria applied to retain only sources with the most reliable measurements. We also describe  the method used to identify non-X-ray galaxies with a significant AGN component.

\subsection{SED modules and parameter space}
\label{appendix_sed}

\begin{table*}
\caption{The models and the values for their free parameters used in CIGALE for the SED fitting.} 
\centering
\setlength{\tabcolsep}{1.mm}
\begin{tabular}{cc}
       \hline
Parameter &  Model/values \\
	\hline
\multicolumn{2}{c}{Star formation history: delayed model and recent burst} \\
Age of the main population & 1500, 2000, 3000, 4000, 5000 Myr \\
e-folding time & 200, 500, 700, 1000, 2000, 3000, 4000, 5000 Myr \\ 
Age of the burst & 50 Myr \\
Burst stellar mass fraction & 0.0, 0.005, 0.01, 0.015, 0.02, 0.05, 0.10, 0.15, 0.18, 0.20 \\
\hline
\multicolumn{2}{c}{Simple Stellar population: Bruzual \& Charlot (2003)} \\
Initial Mass Function & Chabrier (2003)\\
Metallicity & 0.02 (Solar) \\
\hline
\multicolumn{2}{c}{Galactic dust extinction} \\
Dust attenuation law & Charlot \& Fall (2000) law   \\
V-band attenuation $A_V$ & 0.2, 0.3, 0.4, 0.5, 0.6, 0.7, 0.8, 0.9, 1, 1.5, 2, 2.5, 3, 3.5, 4 \\ 
\hline
\multicolumn{2}{c}{Galactic dust emission: Dale et al. (2014)} \\
$\alpha$ slope in $dM_{dust}\propto U^{-\alpha}dU$ & 2.0 \\
\hline
\multicolumn{2}{c}{AGN module: SKIRTOR)} \\
Torus optical depth at 9.7 microns $\tau _{9.7}$ & 3.0, 7.0 \\
Torus density radial parameter p ($\rho \propto r^{-p}e^{-q|cos(\theta)|}$) & 1.0 \\
Torus density angular parameter q ($\rho \propto r^{-p}e^{-q|cos(\theta)|}$) & 1.0 \\
Angle between the equatorial plan and edge of the torus & $40^{\circ}$ \\
Ratio of the maximum to minimum radii of the torus & 20 \\
Viewing angle  & $30^{\circ}\,\,\rm{(type\,\,1)},70^{\circ}\,\,\rm{(type\,\,2)}$ \\
AGN fraction & 0.0, 0.1, 0.2, 0.3, 0.4, 0.5, 0.6, 0.7, 0.8, 0.9, 0.99 \\
Extinction law of polar dust & SMC \\
$E(B-V)$ of polar dust & 0.0, 0.2, 0.4 \\
Temperature of polar dust  & 100 (K)\\
Emissivity of polar dust & 1.6 \\
\hline
\multicolumn{2}{c}{X-ray module} \\
AGN photon index $\Gamma$ & 1.9 \\
Maximum deviation from the $\alpha _{ox}-L_{2500 \AA}$ relation & 0.2 \\
LMXB photon index & 1.56 \\
HMXB photon index & 2.0 \\
\hline
Total number of models (X-ray/reference galaxy catalogue) & 427,680,000/24,552,000 \\
\hline
\label{table_cigale}
\end{tabular}
\tablefoot{For the definition of the various parameters see Sect. \ref{appendix_sed}.}
\end{table*}

The galaxy component is modeled using a delayed star formation history (SFH) model with a function form $\rm SFR\propto t \times exp(-t/\tau)$. A star formation burst is included as a constant ongoing period of star formation lasting 50 Myr, as described by \citep{Malek2018, Buat2019}. Stellar emission is modeled using the single stellar population templates from \cite{Bruzual_Charlot2003} and is attenuated according to the \cite{Charlot_Fall_2000} attenuation law. Nebular emission is modeled using templates based on \cite{VillaVelez2021}. Dust emission, heated by stars, is modeled following \cite{Dale2014}, excluding any AGN contribution. AGN emission is incorporated using the SKIRTOR models from \cite{Stalevski2012, Stalevski2016}. The modules and the parameter space for the SED fitting process is detailed in Table \ref{table_cigale}.

\subsection{SED fitting quality criteria}
\label{appendix_sed_criteria}

To ensure that only sources with the most robust SED fitting measurements are included in our analysis, we applied the criteria used in our previous studies. Specifically, we imposed a reduced \(\chi^2\) threshold of \(\chi^2_r < 5\) \citep[e.g.][]{Masoura2018, Buat2021}. This criterion, excludes 11\% of the X-ray detected sources and 21\% of the non-X-ray galaxies. We also excluded systems for which CIGALE could not constrain the parameters of interest (SFR, M$_*$). To achieve this, we used the two values CIGALE provides for each estimated galaxy property: one corresponding to the best model and the other being the likelihood-weighted mean value (Bayesian). A large difference between these two values indicates a complex likelihood distribution and significant uncertainties. Therefore, we only included sources with \(\frac{1}{5} \leq \frac{SFR_{best}}{SFR_{bayes}} \leq 5\) and \(\frac{1}{5} \leq \frac{M_{*, best}}{M_{*, bayes}} \leq 5\) \citep[e.g.,][]{Mountrichas2022c, Mountrichas2023a, Mountrichas2023b, Mountrichas2023c, Mountrichas2024d} where SFR\(_{best}\) and M\(_{*, best}\) are the best-fit values, and SFR\(_{bayes}\) and M\(_{*, bayes}\) are the Bayesian values estimated by CIGALE. There are 3\,565 X-ray detected AGN and 398\,439 non-AGN sources that meet the aforementioned criteria.

\subsection{Identification of non-X-ray galaxies with strong AGN component}
\label{appendix_SEDAGN}

In the SED fitting analysis, we also model the AGN emission for all sources, independently of whether they are detected in X-rays. This allows us to identify systems with a strong AGN component and exclude them from the non-X-ray detected galaxy sample. We use the AGN fraction parameter, $\rm frac_{AGN}$,  calculated by CIGALE, which is defined as the ratio of the AGN infrared emission ($1-1000\,\mu \rm m$) to the total infrared emission of the galaxy. To identify these systems, we require $\rm frac_{AGN}$ to be higher than 0.2, considering the uncertainties in the measurements, that is $\rm frac_{AGN} - frac_{AGN, err} > 0.2$. A similar criterion has been used in our previous studies \citep[for a more detailed description of the different values that may be applied see][]{Mountrichas2022b}. Approximately 4\% of the non-X-ray detected galaxies are excluded by this criterion.

\section{Mass completeness limits}
\label{appendix_mass_complete}

\begin{table*}[h!]
\centering
\caption{Mass completeness limits in $\log (M_\star/M_\odot)$ in four redshift intervals for the five fields used in our analysis.}
\begin{tabular}{ccccc}
\hline
Field & $\rm 0<z\leq 0.5$ & $\rm 0.5<z\leq 1.0$ & $\rm 1.0<z\leq 1.5$ & $\rm 1.5<z\leq 2.0$  \\
\hline
COSMOS & 8.93 & 9.60 & 9.95 & 10.29  \\
XMM-LSS & 8.97 & 10.04 & 10.50 & 11.12  \\
Stripe82 & 10.17 & 10.83 & 11.24 & 11.56  \\
ELAIS-S1 & 9.03 & 9.94 & 10.41 & 10.91  \\
CDFS-SWIRE & 8.96 & 9.71 & 10.40 & 10.92  \\
\end{tabular}
\label{table_mass_complet}
\end{table*}

To define mass-complete X-ray AGN and reference-galaxy samples, which are essential to avoid potential selection biases in our analysis, we follow the method described in \cite{Pozzetti2010}. This method has also been applied in previous, similar studies \citep[e.g.,][]{Florez2020, Mountrichas2021b, Igo2024}.

We begin by estimating the mass completeness in four redshift bins spanning  \(0 < z < 2\), in intervals of 0.5. For this, we first compute the limiting stellar mass (\(M_{*,lim}\)) of each galaxy using the following expression:

\[
\rm log\,M_{*,lim} = log M_*+0.4(m-m_{lim}),
\]
where $\rm M_*$ is the stellar mass of each source measured by CIGALE, $\rm m$ is the AB magnitude of the source, and $\rm m_{\rm lim}$ is the AB magnitude limit of the survey. This expression estimates the mass the galaxy would have if its apparent magnitude were equal to the survey's limiting magnitude for a specific photometric band. The minimum stellar mass at each redshift interval for which our sample is complete is the 95th percentile of \(\rm log M_{*,lim}\) of the 20\% faintest galaxies in each redshift bin.

We use the \(\rm K_s\) band as the limiting band of the samples, consistent with previous studies \citep{Laigle2016, Florez2020, Mountrichas2021c, Mountrichas2022b}. The \(m_{lim}\) values for each of the five fields used in our analysis are those reported in \cite{Zou2024} (see their Table 1). Using the J or H near-infrared bands does not significantly change the mass completeness limits and does not affect our overall results and conclusions. This holds true even when using a denser redshift grid (\(\rm \Delta z=0.25\)). The mass completeness limits in each redshift bin for the five fields used in our analysis are presented in Table \ref{table_mass_complet}. There are 228\,152 non-X-ray detected galaxies and 3\,046 X-ray AGN that meet the mass completeness limits.

\section{Field-by-field consistency checks}
\label{appendix_fields_separate}

\begin{figure}[h]
\centering
\includegraphics[width=0.45\textwidth, height=7.5cm]{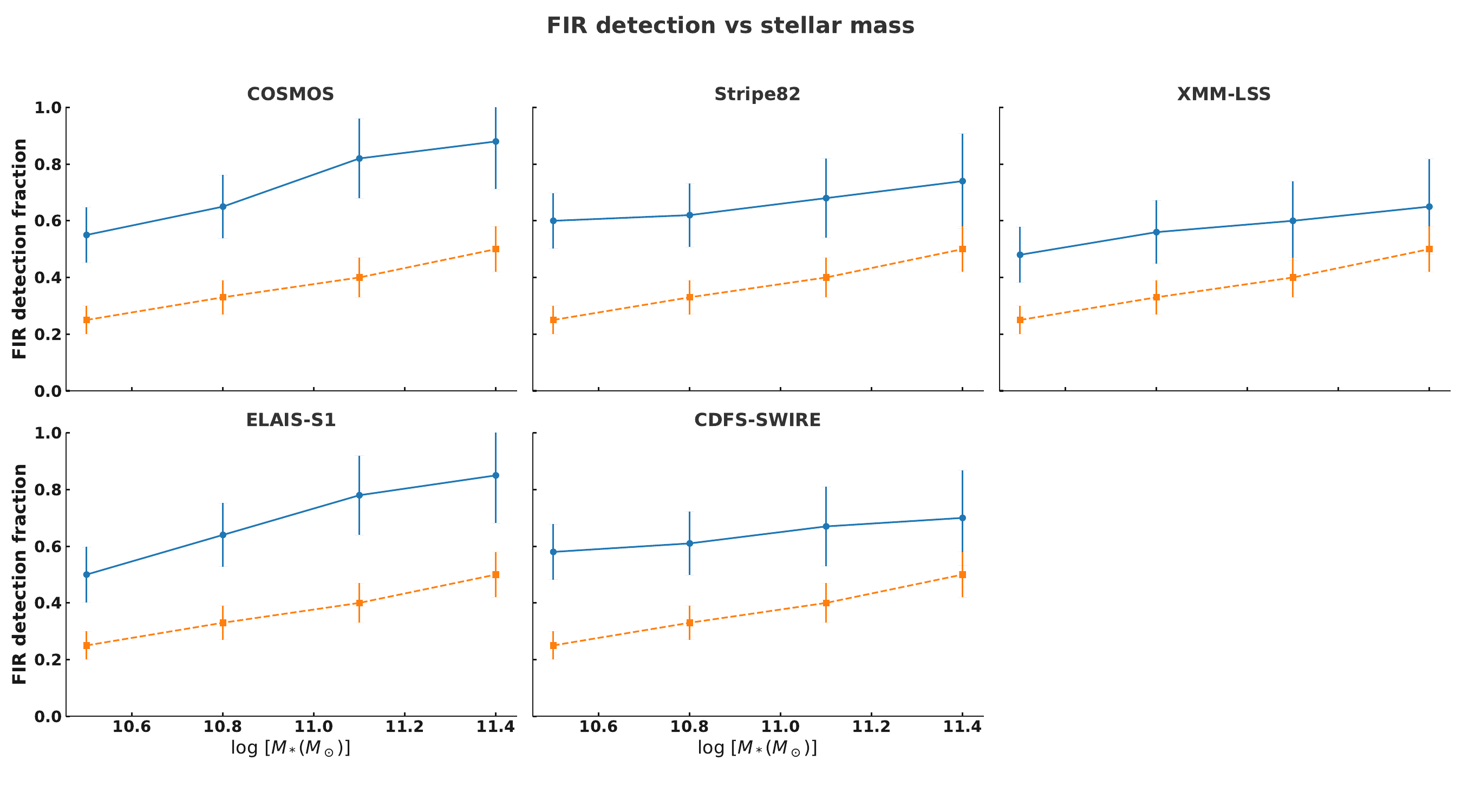}
\caption{FIR detection fraction as a function of stellar mass for X-ray AGN (blue symbols and line) and galaxies (orange symbols and lines) in the five individual fields used in this work, in the redshift range $0.5 \leq z < 1.0$.}
\label{fig:appendix_mstar}
\end{figure}

\begin{figure}[h]
\centering
\includegraphics[width=0.45\textwidth, height=7.5cm]{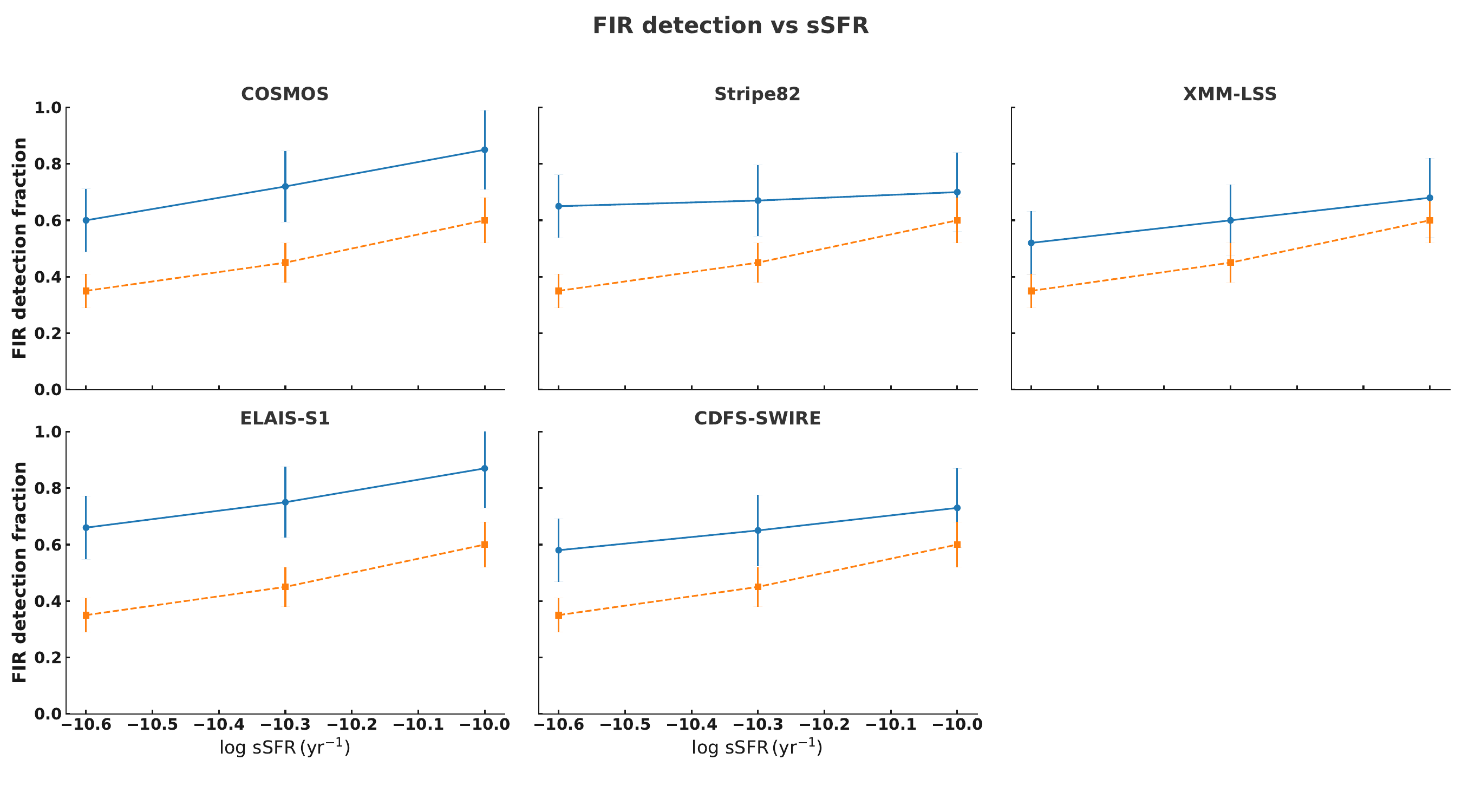}
\caption{Same as Fig. ~\ref{fig:appendix_mstar}, but as a function of specific star formation rate (sSFR).}
\label{fig:appendix_ssfr}
\end{figure}

\begin{figure}[h]
\centering
\includegraphics[width=0.45\textwidth, height=7.5cm]{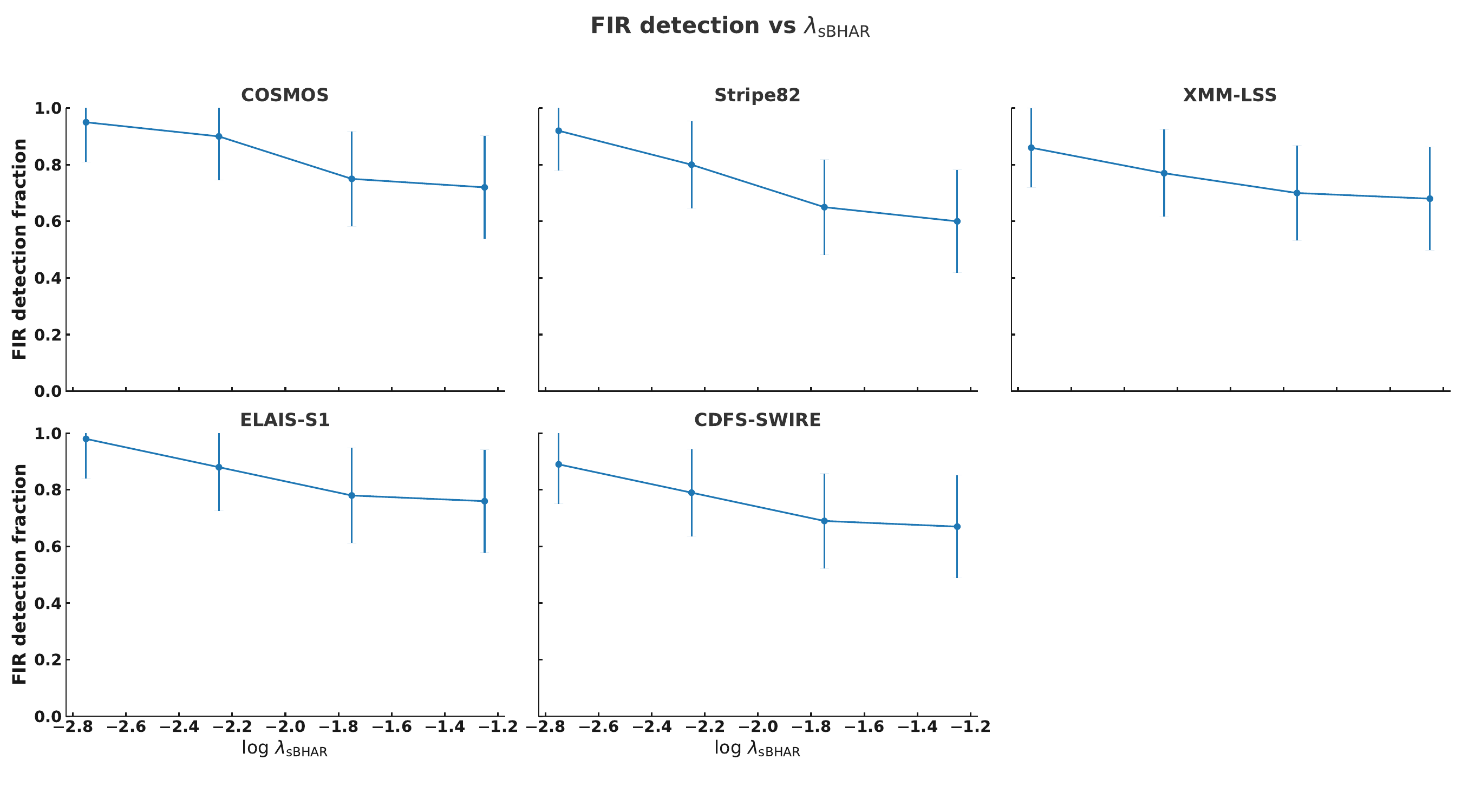}
\caption{Same as Fig. ~\ref{fig:appendix_mstar}, but as a function of specific black hole accretion rate ($\lambda_{\mathrm{sBHAR}}$).}
\label{fig:appendix_lambda}
\end{figure}

To ensure that our main trends are not dominated by any particular field, we repeated the core analysis separately for each of the five fields: COSMOS, Stripe82X, XMM-LSS, ELAIS-S1, and CDFS-SWIRE. In Figures~\ref{fig:appendix_mstar}--\ref{fig:appendix_lambda}, we present the FIR detection fractions for X-ray AGN and non-AGN galaxies as a function of M$_*$, sSFR, and $\lambda_{\mathrm{sBHAR}}$, respectively, in the redshift range $0.5 \leq z < 1.0$. All fields exhibit consistent trends, with only modest variations in normalization or slope, attributable to differences in depth and source sampling.

Although these checks are shown for a single redshift interval for brevity, we confirm that similar consistency is observed across the full redshift range studied in this work. These results demonstrate that our conclusions are robust and not driven by any individual field.

\section{SFR–M$_*$ relation split by far-IR detection}
\label{appendix_sfr_mstar}

\begin{figure}[htbp]
\centering
  \includegraphics[width=0.8\columnwidth, height=5.5cm]{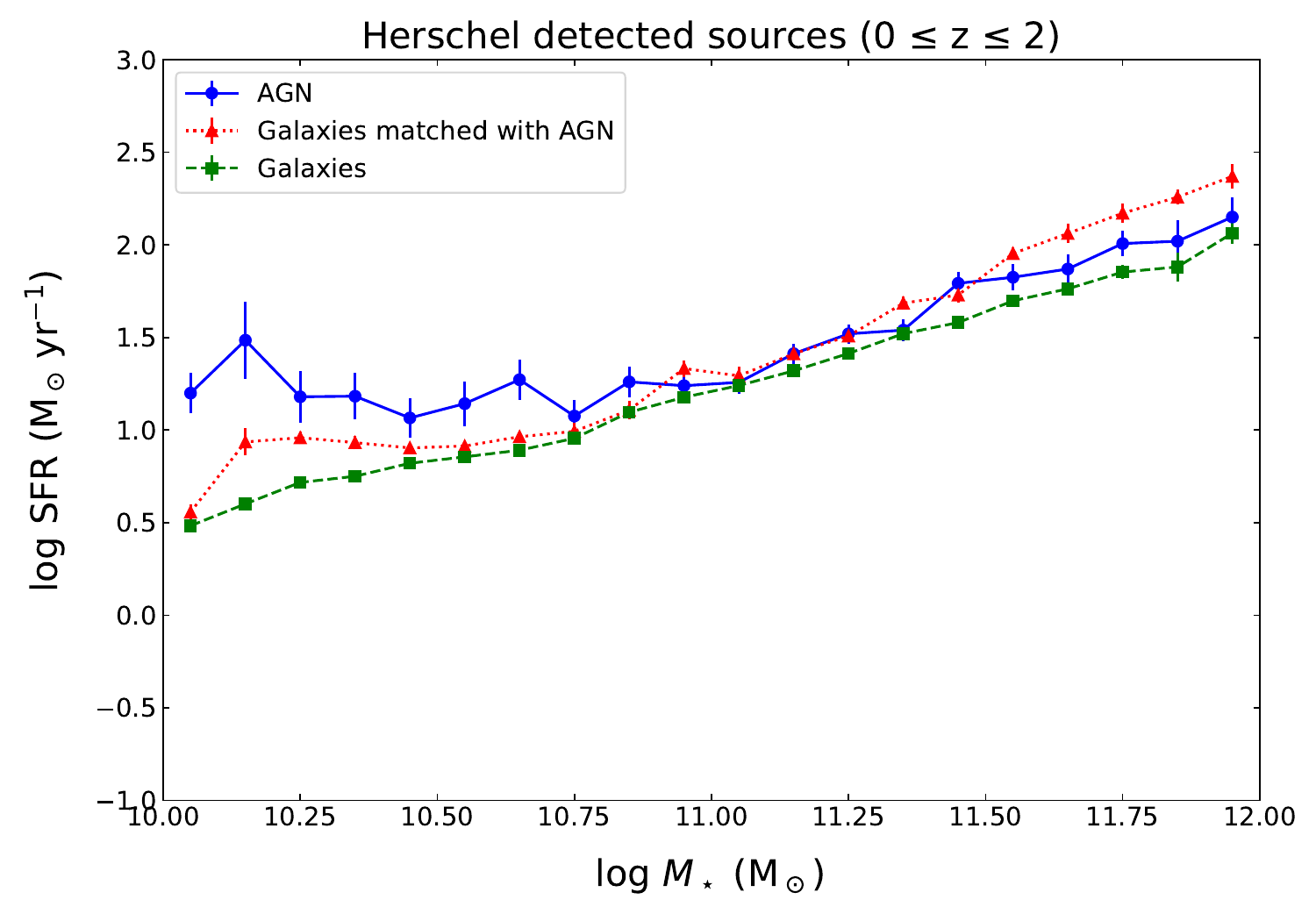}   
  \includegraphics[width=0.8\columnwidth, height=5.5cm]{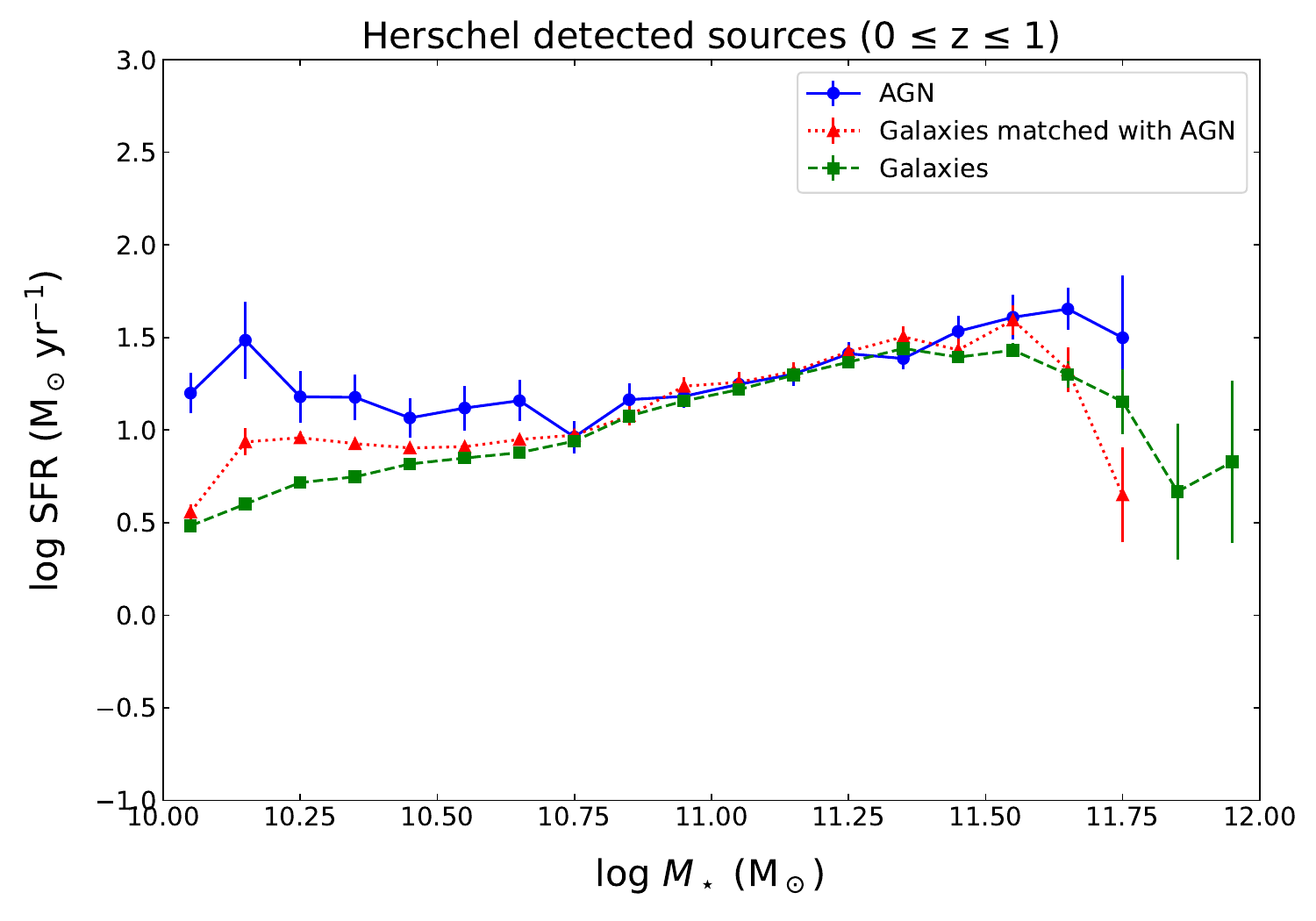} 
  \includegraphics[width=0.8\columnwidth, height=5.5cm]{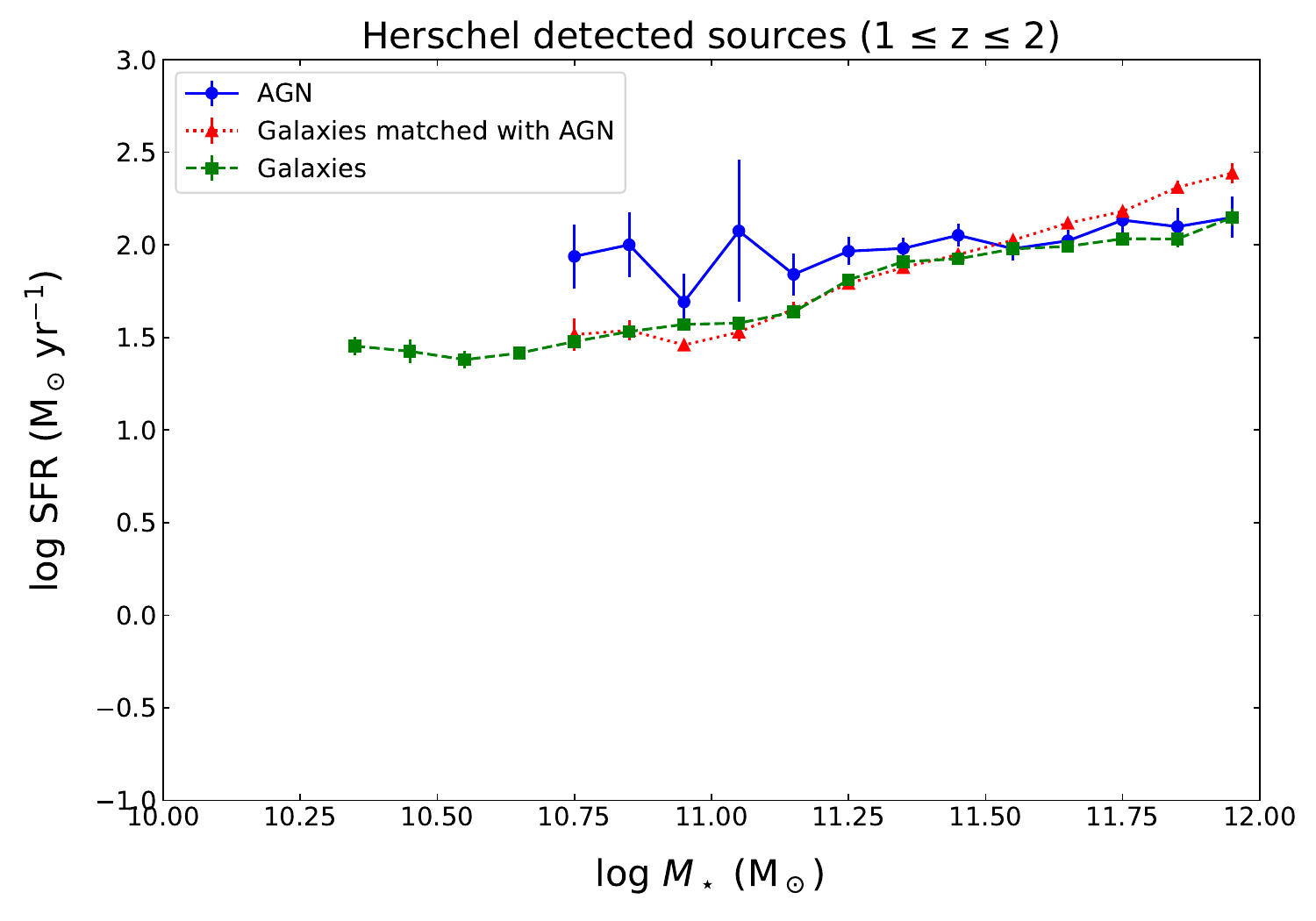} 
  \caption{Log–log relation between SFR and M$_*$ for X-ray AGN, non-AGN galaxies, and AGN-matched control galaxies, using only \textsl{Herschel}-detected sources. (see main text for details). The top panel shows results for the full redshift range, while the middle and bottom panels display the relations at low and high redshift, respectively.}
  \label{fig_sfr_mstar_herschel}
\end{figure}

\begin{figure}[htbp]
\centering
  \includegraphics[width=0.8\columnwidth, height=5.cm]{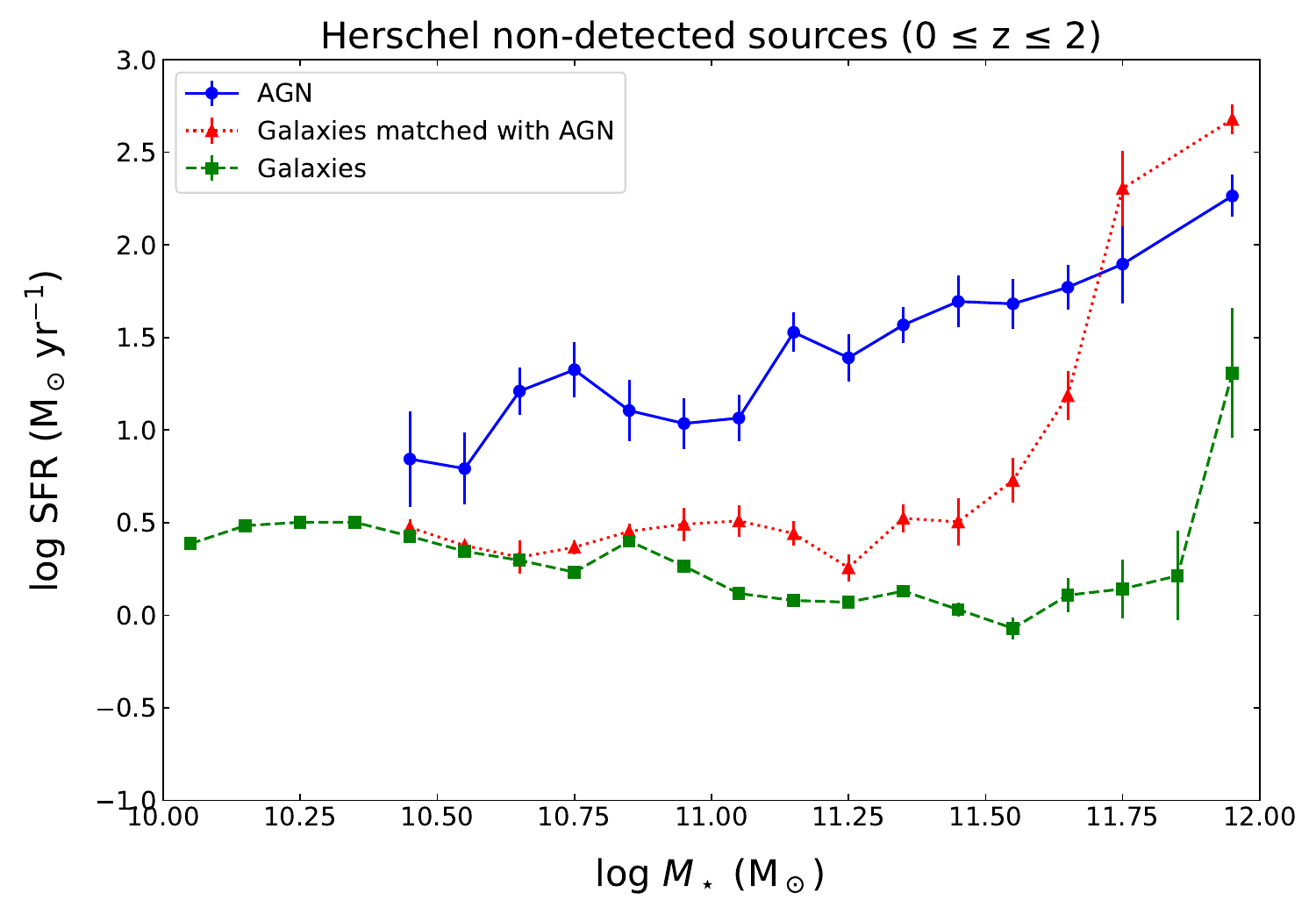}   
  \includegraphics[width=0.8\columnwidth, height=5.cm]{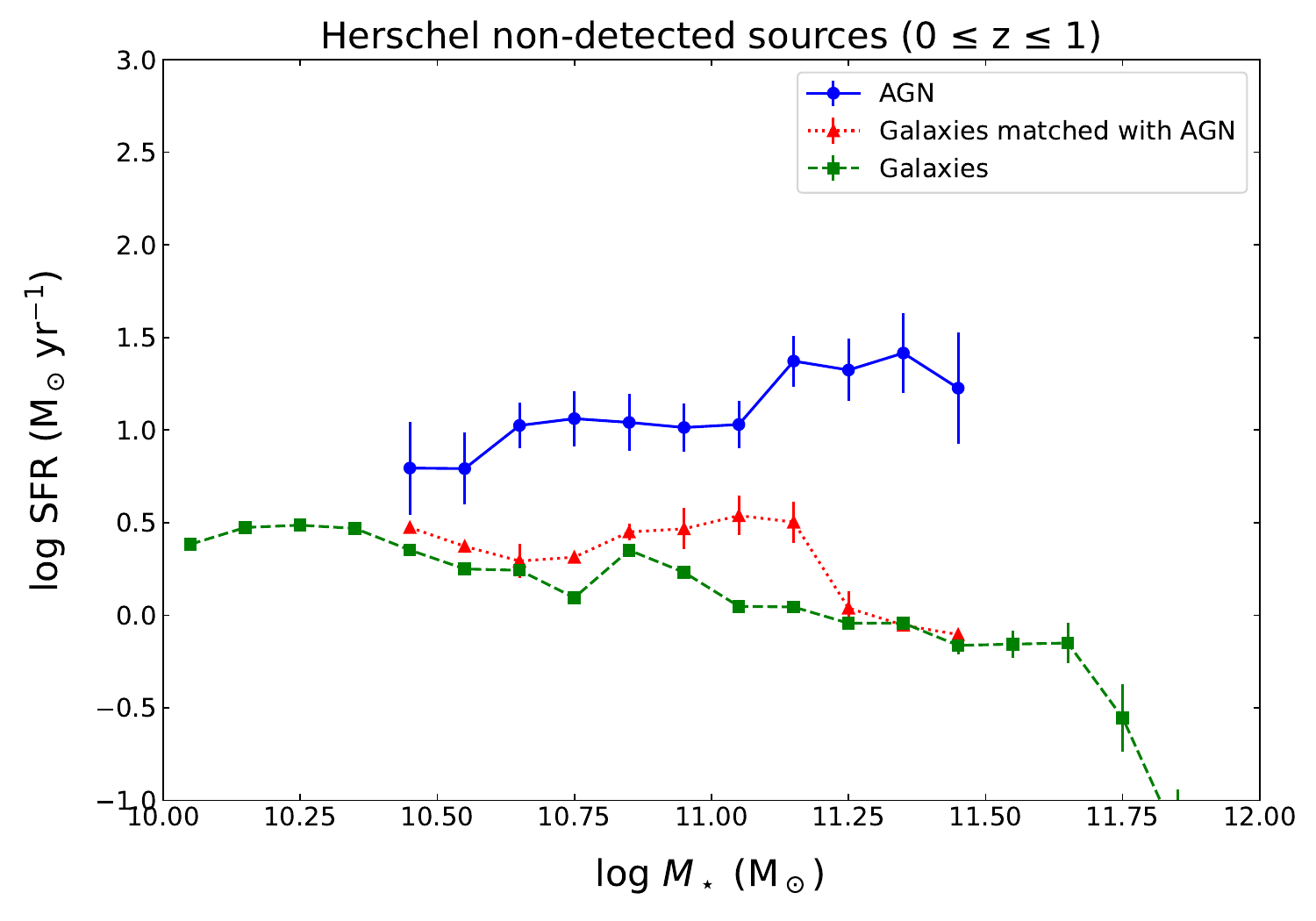} 
  \includegraphics[width=0.8\columnwidth, height=5.cm]{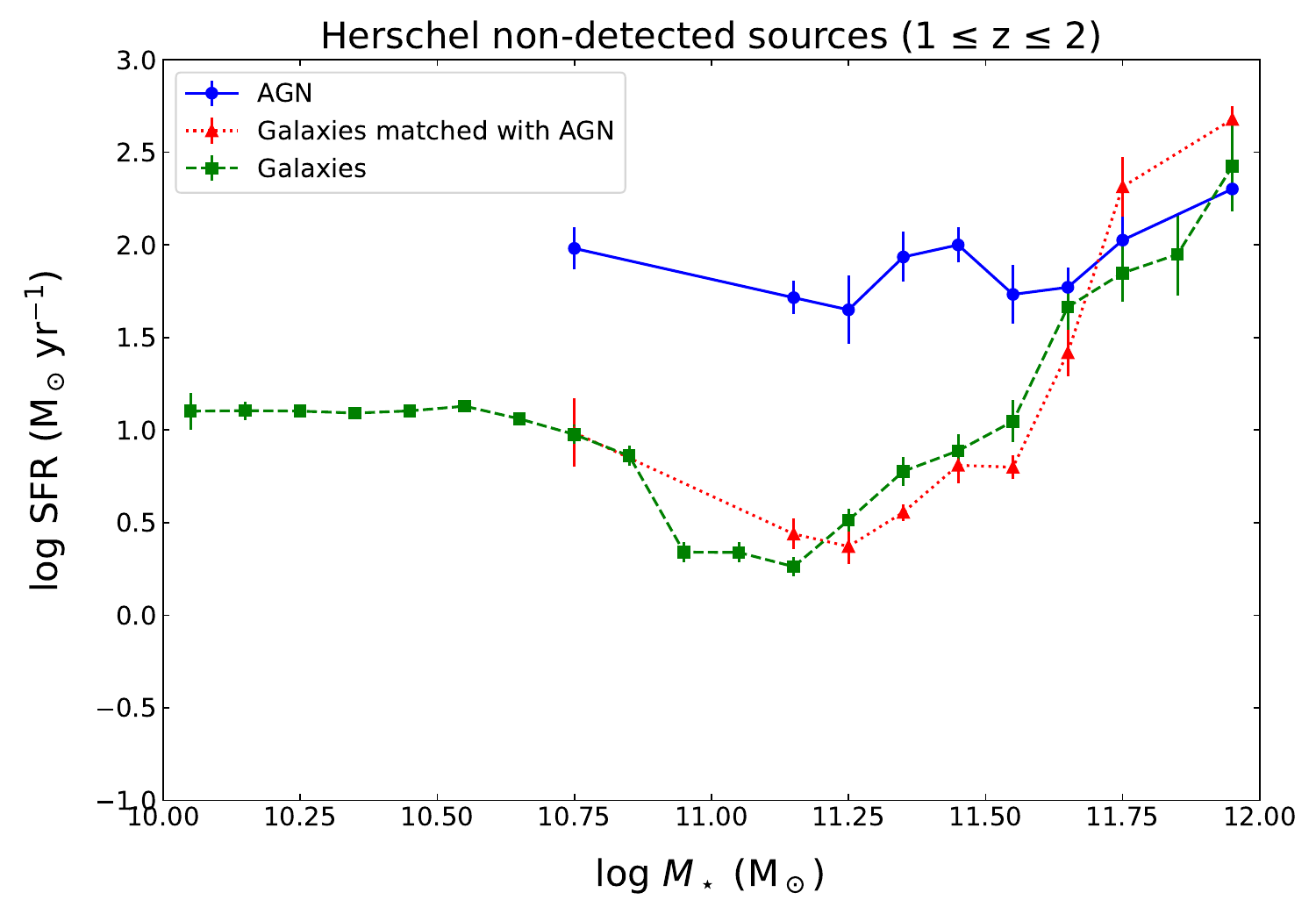} 
  \caption{Same as in Fig. \ref{fig_sfr_mstar_herschel}, but using only sources non-detected by \textsl{Herschel}.}
  \label{fig_sfr_mstar_noH}
\end{figure}

In this section, we present the SFR–M$_*$ relation separately for \textsl{Herschel}-detected and non-detected sources. Fig. ~\ref{fig_sfr_mstar_herschel} shows that among \textsl{Herschel} detected galaxies, AGN and non-AGN systems exhibit similar star formation rates across most stellar masses. In contrast, Fig. ~\ref{fig_sfr_mstar_noH} demonstrates that within the non-\textsl{Herschel} detected sample, AGN host galaxies maintain significantly higher SFR—by approximately 0.5–1 dex—except at the highest stellar masses ($\log[M_*/M_\odot] > 11.5$), where the difference diminishes. These findings support the conclusion that far-IR detection status plays a crucial role in shaping the observed SFR differences between AGN and non-AGN populations.

\section{Classification criteria for AGN based on obscuration}
\label{appendix_agn_obscuration}

For the classification of AGN into type 1 and 2, we use the SED fitting measurements and follow the methodology presented in our previous works \citep[][]{Mountrichas2021b, Mountrichas2024a, Mountrichas2024c}. Specifically, we employ the bayes and best estimates of the inclination, $i$, parameter, derived by CIGALE and classify as type 1 those X-ray AGN with $i_{best}=30^{\circ}$ and $i_{bayes}<40^{\circ}$, while type 2 sources are those with $i_{best}=70^{\circ}$ and $i_{bayes}>60^{\circ}$.

\cite{Mountrichas2021b}, showed that application of the above criteria yielded an accuracy of approximately 85\% in classifying type 1 AGN. A similar level of accuracy was observed for the completeness of type 1 source identification. However, for type 2 sources, CIGALE's performance was approximately 50\%, both in terms of reliability and completeness. The reliability is defined as the fraction of the number of type 1 (or type 2) sources classified by the SED fitting that are similarly classified by optical spectra. The completeness refers to how many sources classified as type 1 (or type 2) based on optical spectroscopy were identified as such by the SED fitting results. For the purposes of our current study, our primary focus is on evaluating the reliability performance of CIGALE. In \cite{Mountrichas2021b}, it was shown that the majority ($\sim 82\%$) of the misclassified type 2 sources exhibit elevated polar dust values (E$_{B-V}>0.15$; refer to their Fig. 8 and Section 5.1.1). Consequently, we follow our previous works and exclude these sources from our analysis and categorized as type 2 those AGN that meet the specified inclination angle criteria and also possess polar dust values lower than E$_{B-V}<0.15$. Application of the aforementioned criteria results in 1\,690 type 1 and 212 type 2 X-ray AGN.

\section{Classification criteria for AGN based on X-ray absorption}
\label{appendix_agn_absorption}

\begin{figure}
\centering
  \includegraphics[width=0.85\columnwidth, height=5.cm]{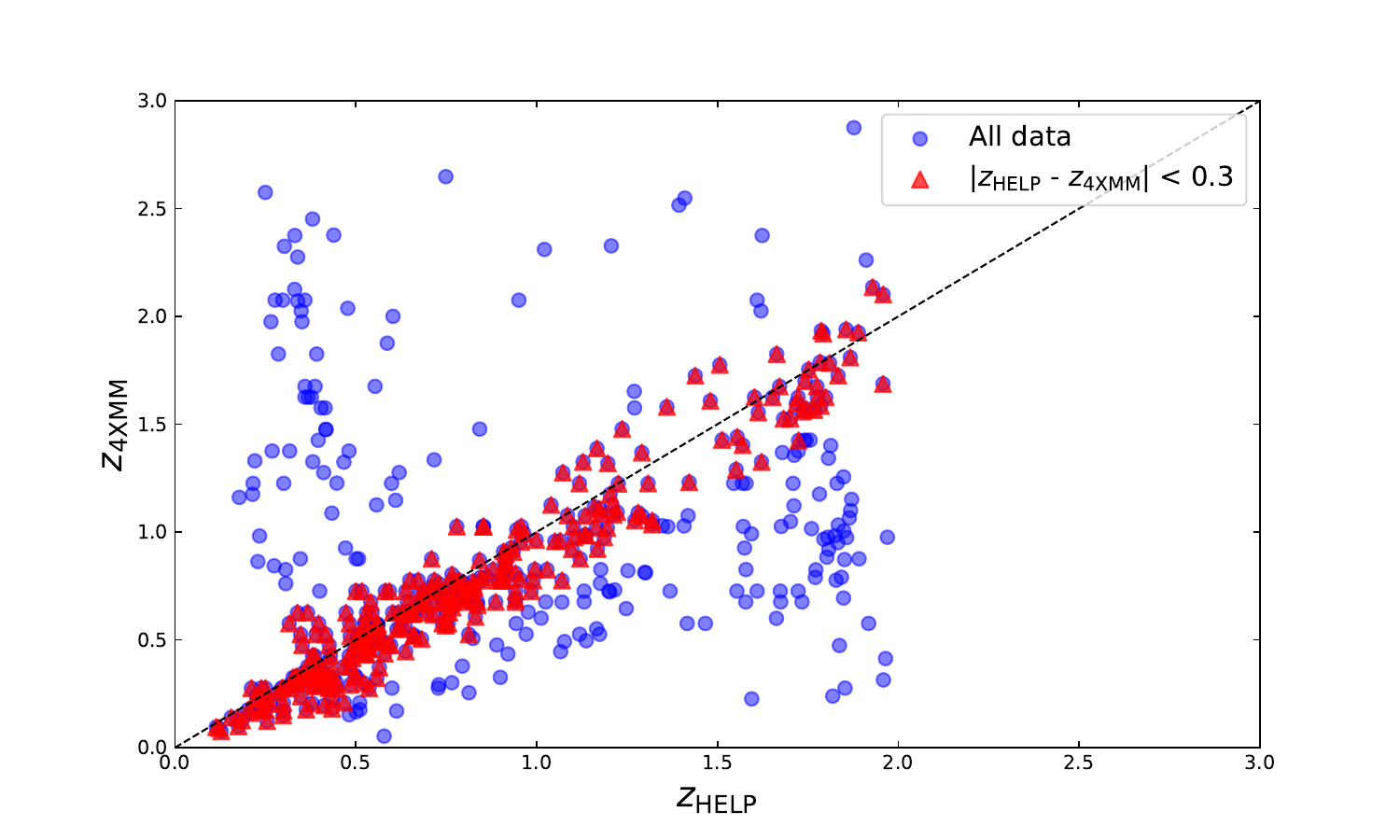} 
  \caption{Comparison of the two redshift calculations available for the 484 X-ray AGN (blue circles), that are common between our X-ray dataset described in Sect. \ref{sec_final_samples} ($\rm z_{HELP}$) and the 4XMM catalogue ($\rm z_{4XMM}$). Sources that meet the redshift difference criterion are shown by red triangles (see text for more details).}
  \label{fig_redz_comp}
\end{figure}

To examine the properties of X-ray absorbed and unabsorbed AGN as a function of their incidence in the far-IR, we matched the X-ray dataset from our previous analysis with the 4XMM-DR11 catalogue \citep{Webb2020}. This catalogue which provides measurements of X-ray spectral parameters within the framework of the XMM2Athena{\footnote{http://xmm-ssc.irap.omp.eu/xmm2athena/}} project. The 4XMM-DR11 is the fourth-generation catalogue of serendipitous X-ray sources from the European Space Agency's (ESA) XMM-Newton observatory, created by the XMM-Newton Survey Science Centre (SSC) on behalf of ESA. This catalogue contains 210,444 sources with available spectra from one or more detections. The classification and fitting processes for these sources are detailed in \cite{Viitanen2025} and in Section 2 of \cite{Mountrichas2024b}. The catalogue includes 35,538 AGN with available redshift information, of which 8,467 have spectroscopic redshifts. Photometric redshifts for the remaining sources are calculated using the methodology described in \cite{Ruiz2018}. For the X-ray spectral analysis of these AGN, a Bayesian spectral fit was performed using the \texttt{BXA} tool \citep{Buchner2014}, which connects \texttt{XSPEC} with the nested sampling package \texttt{UltraNest} \citep{Buchner2021}. The fitting was carried out using an absorbed powerlaw model \citep{Viitanen2025}

To incorporate the hydrogen column density parameter (N$_H$) from the X-ray spectral analysis, we cross-matched the 2\,417 X-ray AGN used in our analysis (see Section \ref{sec_final_samples}) with the 35,538 AGN in the 4XMM dataset, finding 584 common AGN. These 584 X-ray AGN have two redshift measurements, either spectroscopic or photometric: one from the HELP catalogues used for SED fitting measurements, and one from the 4XMM catalogue used for X-ray spectral parameters. To assess the consistency of these two redshift measurements, we use two widely adopted statistical parameters: the normalized absolute median deviation, $\sigma_{\text{nmad}}$, and the percentage of outliers, $\eta$. 

$\sigma_{\text{nmad}}$ is defined as:
\[
\Delta(z_{\text{norm}}) = \frac{z_{\text{HELP}} - z_{\text{4XMM}}}{1 + z_{\text{HELP}}},
\]
\[
\text{MAD}(\Delta(z_{\text{norm}})) = \text{Median}(|\Delta(z_{\text{norm}})|),
\]
\[
\sigma_{\text{nmad}} = 1.4826 \times \text{MAD}(\Delta(z_{\text{norm}})). \quad (1)
\]

The percentage of outliers, $\eta$, is defined as:
\[
\eta = \frac{100}{N} \times (\text{Number of sources with } |\Delta(z_{\text{norm}})| > 0.15).
\]

We find $\eta = 35.7\%$ and $\sigma_{\text{nmad}}=0.11$. To ensure only sources with consistent redshift measurements are included, we apply a criterion requiring the difference between the two redshift calculations to be less than 0.3, resulting in $\sigma_{\text{nmad}}=0.05$ and $\eta = 3.3\%$, and yielding 346 X-ray AGN. Notably, varying this criterion within the range of 0.1-0.5 does not affect our overall results and conclusions, but changes the size of the available X-ray sample. Fig. \ref{fig_redz_comp} shows the comparison of the two redshift calculations. Out of these 346 X-ray AGN, 289 are within the stellar mass range used in our analysis ($\rm 10<log\,[M_*(M_\odot)]<12$).

We use an $N_H = 10^{22}\ (\text{cm}^{-2})$ threshold to classify the 289 X-ray AGN into X-ray absorbed (high \(N_H\)) and X-ray unabsorbed (low \(N_H\)). Out of these, 38 sources are classified as X-ray absorbed, with 33 of them detected by \textsl{Herschel} (approximately 87 percent). Meanwhile, 251 sources are classified as X-ray unabsorbed, with 163 detected by \textsl{Herschel} (approximately 65 percent).

\end{document}